# The Andromeda Study: A Femto-Spacecraft Mission to Alpha Centauri

Andreas M. Hein, Kelvin F. Long, Dan Fries, Nikolaos Perakis, Angelo Genovese, Stefan Zeidler, Martin Langer, Richard Osborne, Rob Swinney, John Davies, Bill Cress, Marc Casson, Adrian Mann, Rachel Armstrong

Andreas.Hein@i4is.org
Initiative for Interstellar Studies, 27/29 South Lambeth Road, London, SW8 1SZ

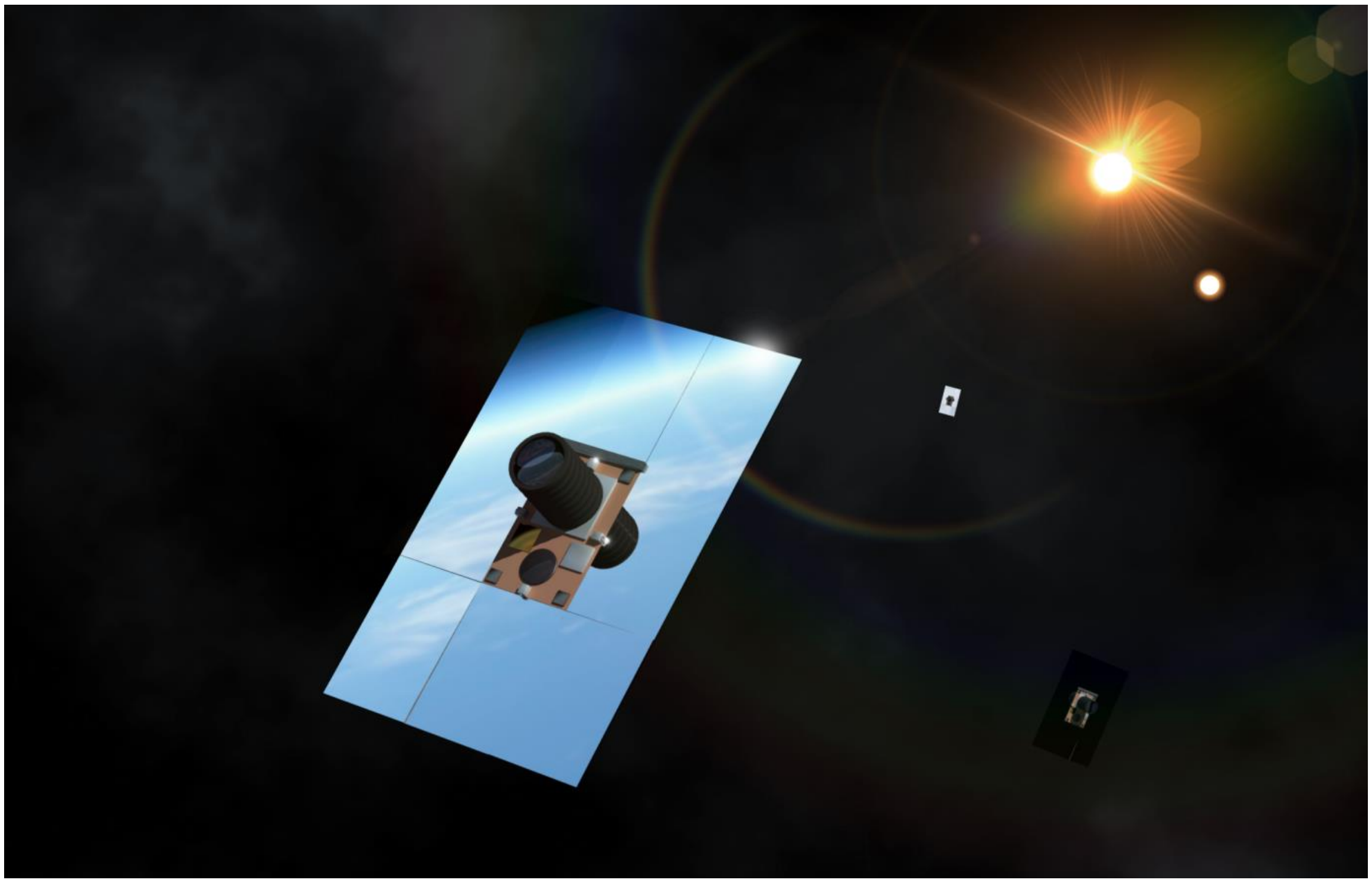

## Abstract

This paper discusses the physics, engineering and mission architecture relating to a gram-sized interstellar probe propelled by a laser beam. The objectives are to design a fly-by mission to Alpha Centauri with a total mission duration of 50 years travelling at a cruise speed of 0.1c. Furthermore, optical data from the target star system is to be obtained and sent back to the Solar system. The main challenges of such a mission are presented and possible solutions proposed. The results show that by extrapolating from currently existing technology, such a mission would be feasible. The total mass of the proposed spacecraft is 23g and the space-based laser infrastructure has a beam power output of 15GW. Further exploration of the laser – spacecraft trade-space and associated technologies are necessary.

# 1. INTRODUCTION

Concepts for beam-propelled interstellar missions have been proposed notably by [1] for a microwave-propelled gram-sized probe and [2] for a laser-propelled fly-by probe with a 333kg payload, propelled by a 65GW space-based laser infrastructure. The spacecraft would be accelerated over a distance of 0.17 light years. More recent laser sail concepts such as [3], [4] and [5] have proposed the use of dielectric materials, which have a reflectivity larger than 99% and melting temperatures of between 2000-3000K for the laser sail. Furthermore, they propose much higher accelerations of hundreds to thousands of g that are enabled by the higher temperature limit of the sail material. Besides dielectric materials, [6]–[8] have proposed the use of Graphene-based sails that have a melting temperature above 4000K and a very low density. However, one drawback is the low reflectivity of the material. Although [1] has proposed a probe with a total mass of one gram, such low masses seemed to be infeasible by that time. However, recent developments in micro-electronics has lead to the emergence of small spacecraft such as CubeSats [9] with masses between 1 to a dozen kg, FemtoSats with masses below 100g [10], and AttoSats with masses below 10g such as ChipSats [11]. CubeSats have extended their applications from educational purposes to scientific and industrial applications. Recently, interplanetary CubeSats have been proposed as a means for low-cost exploration of outer space [12]. Landis [13] has already explored the prospects of decreasing the payload mass for a laser-propelled fly-by probe in 1995 by extrapolating from mass decreases at that time. He argues that a decrease to 8kg might be feasible by 2030. One recent proposal [5] has explored the trade-space for gram-sized interstellar probes. However, details regarding the subsystems and the overall architecture of spacecraft and the space-based laser infrastructure have been left open.

*Extrapolating from these trends, this paper addresses the question, how far a FemtoSat-scale interstellar probe would be feasible, exploring a variety of technology options and proposing a baseline concept.*

The problem is defined to address the scenario of a 10-100 gram interstellar probe sent to the Alpha Centauri star system in a total mission time of 50 years, travelling at a cruise speed of 0.1c. Data is to be obtained at the target system and returned back to the solar system. The propulsion system is to be laser based. A set of questions were tasked, of which this brief note sets out multiple solutions. It is necessary to first scope the problem.

The distance to the Alpha Centauri system is 4.3 Light Years, or approximately 272,000 AU. Travelling at 0.1c (30,000 km/s) over this distance would take 43 years, allowing for up to 7 years for the acceleration phase, implying a lower acceleration bound of 0.136 m/s$^2$. Table 1 below shows several acceleration scenarios for a 100 gram probe assuming a reflectivity of unity, and the distances obtained end of boost assuming that 0.1c is the cruise target speed. These calculations are non-relativistic but are presented as a starting point for the analysis.

For this brief study it was necessary to also make some basic assumptions in the absence of information. These are briefly defined:

1. Flyby mission only (no deceleration)
2. Launch date 2025-2035.
3. No consideration is given for launch costs, which are considered a separate matter.

*Table 1-1: Example Probe Scenarios for an interstellar mission (100 grams)*

| Acceleration (m/s$^2$) | Power (MW) | Boost duration (years) | End of Boost distance (AU) |
|---|---|---|---|
| 0.136 | 2.04 | 7 (2,555 days) | 22,230.9 |
| 0.190 | 2.85 | 5 (1,825 days) | 15,788.7 |
| 0.317 | 4.75 | 3 (1,095 days) | 9,483.2 |
| 0.951 | 14.26 | 1 (365 days) | 3,161.0 |
| 1.903 | 28.53 | 0.5 (183 days) | 1,581.4 |
| 3.805 | 57.07 | 0.25 (91 days) | 790.5 |
| 9.513 (~1g) | 142.69 | 0.1 (37 days) | 316.2 |
| 19.026 | 285.39 | 0.05 (18 days) | 158.1 |
| 38.052 | 570.78 | 0.025 (9 days) | 79.05 |

It needs to be acknowledged straight up that a 50 year mission to the stars is extremely challenging whatever the propulsion system being utilised and many example studies to address this are well documented [14]. This would become even more so if it was also desired to effect some form of deceleration, although it is assumed not to be the case for this study. Yet, at 0.1c the encounter time at the target will be days. If we assume a typical 100 AU stellar system diameter, then at 0.1c the probe would have passed through the entire system in 0.0158 years or around 6 days. For this reason, it might be worthwhile for any future iteration of this work to look at ways of slowing the probe down as it approaches the target.

Forward [15] has examined the idea of decelerating by electrically charging the probe to turn using the Lorentz force in the interstellar magnetic field [16], reversing the sail velocity so that the laser can decelerate it. He concludes that there is sufficient doubt about the strength of the interstellar magnetic field that we do not know if this is possible.

Andrews and Zubrin [17] discuss several difficulties relating to maintaining a focused beam on the sail over interstellar distances. They conclude that for the two-stage LightSail, the mirror must maintain surface quality to a non-plausible tolerance as it decelerates the second stage. They also discuss the fact that the laser dispersion due to beam quality, jitter, etc. must be considerably better than beam qualities of existing state of the art beams. Hence, they propose braking interstellar ships using drag from the interstellar medium or the solar wind by using magnetic sails. More recently, the concept has been evaluated with respect to laser-propelled interstellar missions [18], [19]. Perakis and Hein [19] propose combining a magnetic and electric sail for quicker deceleration than each of the sails alone.

Another possible option to effect some deceleration of the sail is to provide for a central concentric ring in the sail, enough for gasses to be ejected, and then to use the incoming particle stream hitting the sail as an energy source for a photovoltaic mini-electric (ion) engine. Such a concept was described by Landis [4]. It is not likely that this would provide sufficient thrust to decelerate to orbital velocity but the combination of the sail pressure and ion drive may be enough to increase the encounter time from days to weeks.

In the remainder of this paper we will lay out the various technical issues that are relevant to the probe design and therefore set out the necessary program of work that needs to be conducted in order to make a flight feasible. Where possible we provide more than one solution to some of the technical problems so as to provide options and inform the design reference point.

# 2. SPECIFIC TECHNICAL CHALLENGES

Several technical challenges are briefly addressed. Some additional issues are covered for completion.

## 2.1 Deep Space Navigation

It is expected that due to the small size of the probe it will have limited ability to cause directional changes to its trajectory after the acceleration phase. It is possible to eject gas so as to change the pointing but this will have little effect compared to the directional velocity of the vehicle travelling at 0.1c. If it were possible however, with a larger probe mass, then navigation by pulsars and the known brightest stars within the stellar neighbourhood would be key. For pulsar navigation, a miniaturized X-ray telescope would be required, which are under development for CubeSat missions today [20]–[22]. Another possibility is to use miniaturized star trackers with star maps for various locations along the trajectory [23], [24].

Another, passive approach is to use the orientation of a laser beam locked on to spacecraft or orientation of incoming photons from the communication system in conjunction with the red-shift of incoming photons from the communication system or from the reflected photons of the beam-lock to calculate the distance. The precision of this approach is limited by the precision of red-shift measurements (current equipment has errors <1%) [25] and the laser beam diameter at the position of the probe. The best case is thus probe arrival at Alpha Centauri. Initial considerations lead to the following results:

The distance can be determined very precisely, while the position within the beam spot is known with a confidence of up to +-0.004% of an AU. Moreover, the probe needs to be equipped with very accurate pointing capabilities, but not only for navigation. When passing through the target system at 0.1c very high tracking and pointing accuracy is required as well, otherwise any images will be useless, especially when trying to approach an object more closely. The LISA targeting system is very large for what we are trying to achieve, but it uses technologies that can be miniaturized and achieves nano-radian accuracy. The James Webb space telescope also achieves nrad pointing (~24 nrad), so that there is interest and progress in this field. Miniaturized gyros and accelerometers (MEMS) to aid in the pointing motion and measurement of the orientation are a requirement too. Due to the delay in receiving signals on both ends the position of the spacecraft will be merely predicted at some point and deviations from the nominal trajectory can hardly be corrected in real time, requiring almost full autonomy of the spacecraft or group of spacecraft.

## 2.2 Communications between the Probe and Earth

For the communications systems, we look at several options. This includes radiofrequency and optical communication. We consider the following communication architectures:

- Radiofrequency (RF) communication via antenna on spacecraft and receiver in Solar system
- Laser communication via laser on spacecraft and receiver telescope in Solar system
- Radiofrequency communication via the Sun's gravitational lensing effect.
- Optical occultation communication as back-up solution: Opto-electrical parts of sail occult the target star. Changing occultation results in a "morse-code" signal.
- Trailing communication: Trailing spacecraft are launched that serve as relay stations. Either radiofrequency or laser systems are used in conjunction.

As an example for communication requirements, we look at the Robert Forward Starwisp paper [1] that assumes $8 \cdot 10^6$ bits for an image taken by the spacecraft with a resolution of 1000x1000 pixel (8 bit per pixel for 256 shades of grey). At 80 bits per hour, it would take 100 hours to finish transmission of one file on the spacecraft side. Considering the communications delay at interstellar distances (several lightyears) this might not be an issue, however, sufficient power must be available and the downlink cannot be interrupted during transmission and receiving operations, respectively.

Messerschmitt [26] considered the theoretical limits of low power information transmission in the radiofrequency regime. The author argues that contrary to terrestrial and near-earth communication where bandwidth availability is limited, interstellar communication is energy limited and the underlying equations suggest that maximizing the utilized signal bandwidth is beneficial. Considering the limiting case of an infinite bandwidth, he calculated 8 photons per 0.46 Wh as the fundamental energy limit, i.e. with 4.6 W 80 photons per hour can be sent. The fundamental limit for received energy per bit is calculated as $7.66 \cdot 10^{-23}$ J/bit (page 227 of Messerschmidt's thesis), assuming a total noise temperature of 8 K.

For our own link budget calculations we make the following common assumptions: a) the maximum distance is 4.3 lightyears (Alpha Centauri); b) the output power of transmitter is 1W; c) the bandwidth is 1 Mhz, which can be seen as representative for a large bandwidth RF sytem and a typical laser linewidth; d) the receiver antenna is 70 m in diameter, which is equivalent to the currently largest antennas in the DSN.

Moreover, we assume an antenna efficiency of 65% with the gain profile of pencil beam antennas, no channel coding and the downlink as the worst-case scenario since it is reasonable that more power is available on the receiver side in the solar system. Constant improvements and developments in the communications sector are likely to only improve upon our estimates.

The resulting RF link budget assuming K-band utilization is shown in Figure 1. Since we want to miniaturize the spacecraft, thus, limiting the available transmitter power, transmitter antennas have to approach diameters of hundreds of meters to allow for the transmission of just 1 bit/s. Even larger antennas are necessary if positive Signal-to-Noise ratios are required to separate the signal from background noise. A solution could be an ultrathin foldable antenna, based on graphene sandwich material, similar to the material used for the laser sail. Satellite antennas with diameters of close to 20m in diameter have already been flown on geostationary satellites, for example, on the ETS-VIII mission. Hence, a 100m antenna is deemed to be feasible by extrapolating existing technology.

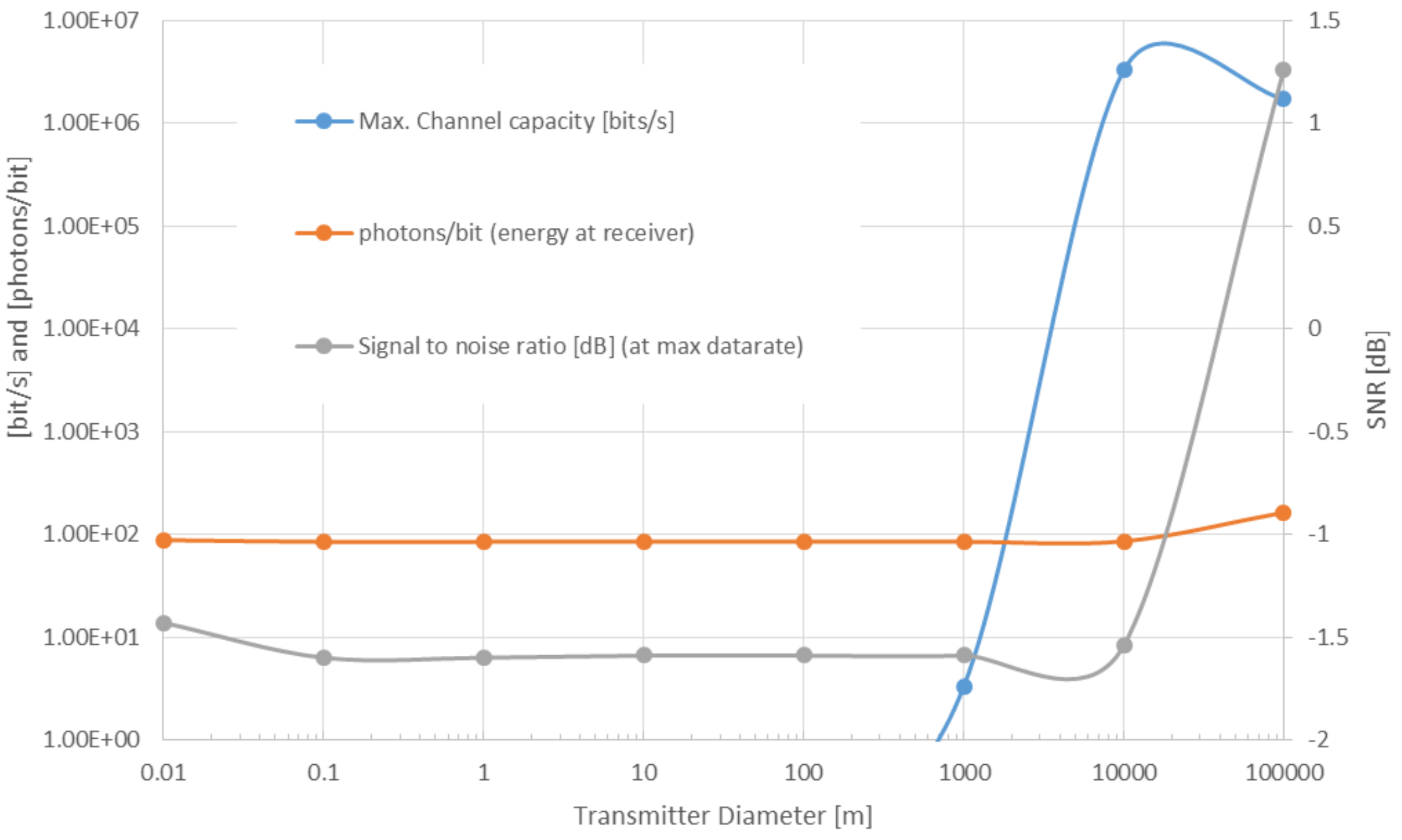

*Figure 1: RF link budget calculations at the fundamental limit*

Optical communication is another option, where lasers are used instead of radiofrequency antennas. The higher frequency of lasers allows for a higher data rate. However, since lasers are highly directional, accurate pointing of the beam is required to avoid excessive losses. Lasers for interstellar probe communication have been proposed by [8] and Lubin [5]. [5] proposes to submit the data via short laser power "bursts" to achieve high data rates over short periods of time. Moreover, the development for chip sized photonic phased laser arrays is actively pursued [27]. For our optical link budget at 500 nm (green) wavelength laser is considered. We assume a pointing accuracy of at least 1 nrad, which is about an order of magnitude better than what currently seems feasible [28][29]. Thus, it is within the realm of possibilities considering future developments in the area. A Gaussian beam with a divergence of 6.1 μrad is assumed, the path loss is mainly due to spill over the receiver while pointing loss is based on pointing accuracy and an estimated standard deviation[1]. We further postulate that ≥1 bit/photon data encoding is possible since different methods working in this direction have already been presented [30][31]. With a 1 cm laser beam transmitter on the spacecraft side the resulting link budget is shown in Figure 2.

[1] Dolinar et al., "Fundamentals of Free-Space Optical Communication", (2012)

Losses pertaining to the optical communication system that have not been considered are related to optical emissions from the telescope/ array, emissions from our solar system dust both scattering sunlight and emitting thermal radiation and the Cosmic Infrared Background [5]. For terrestrial receivers the atmospheric scattering and emissions present further complications.

To give an impression of the RF and optical antenna sizes needed for a specific downlink speed Figure 3 and Figure 4 are presented. It gives the achievable data rates at a distance of 4.3 light years and a transmitter power of 1W as a function of both transmitter and receiver antenna diameter assuming 65% antenna efficiency. At first glance, the plots suggest that optical communications are far superior in terms of achievable datarate for a given setup. However, assumptions have been made regarding pointing accuracy, jitter and bit/photon capacity that might be too optimistic, and, as mentioned above, some losses have not been considered. Thus, more detailed analysis is required to establish a truly reliable optical communications trade space.

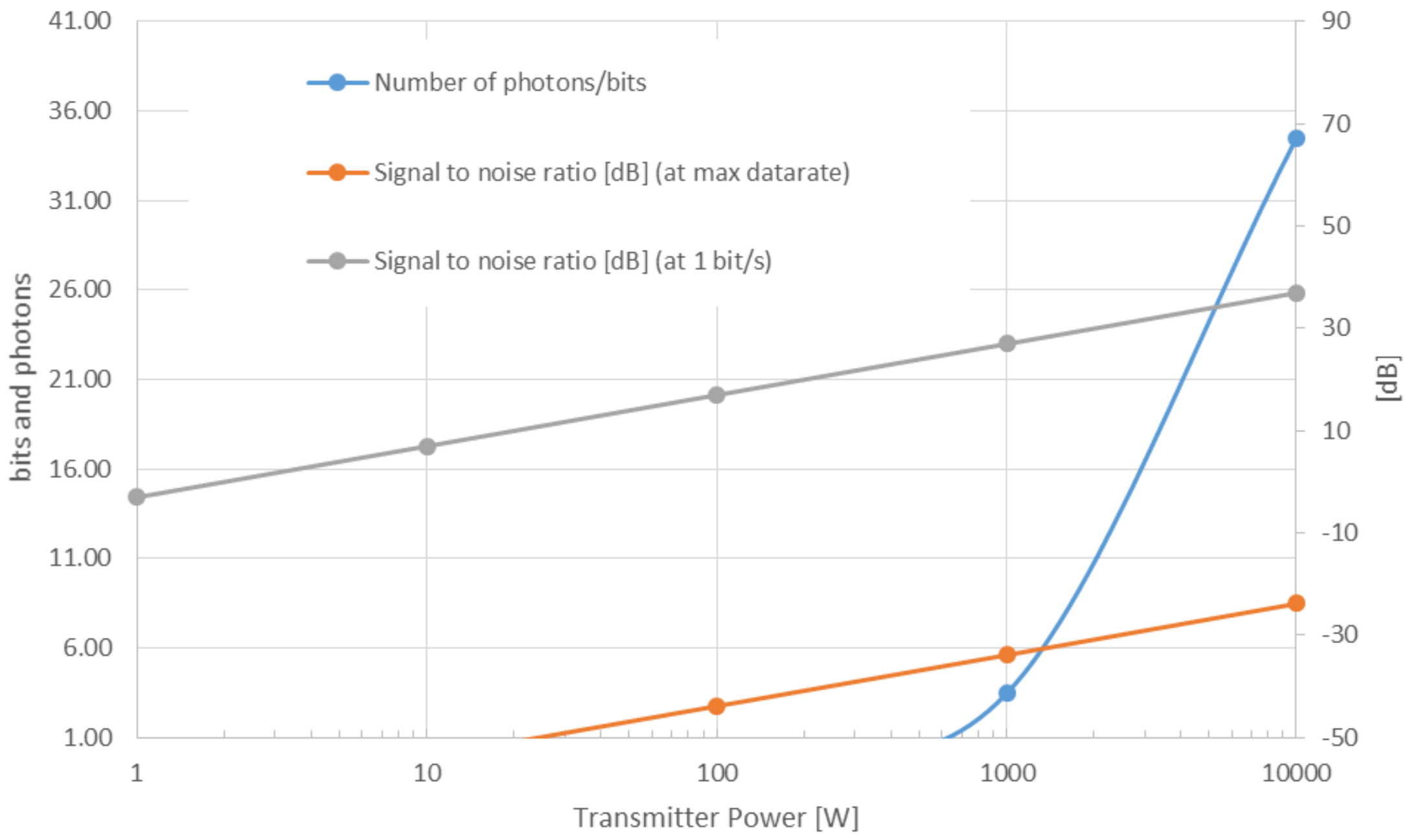

*Figure 2: Optical link budget calculations*

To improve the gain at the receiver end [32], the Sun's gravitational lens could be used as a gigantic receiver station, as Drake discovered [33]. The electromagnetic signals are amplified by the Sun's gravitational field and a receiver spacecraft at the Sun's focal point 550 AU away can pick up the signal [34]. Technical challenges are to position the spacecraft at a straight line with the Sun and the interstellar probe over these distances and to keep track of the signal from the probe.

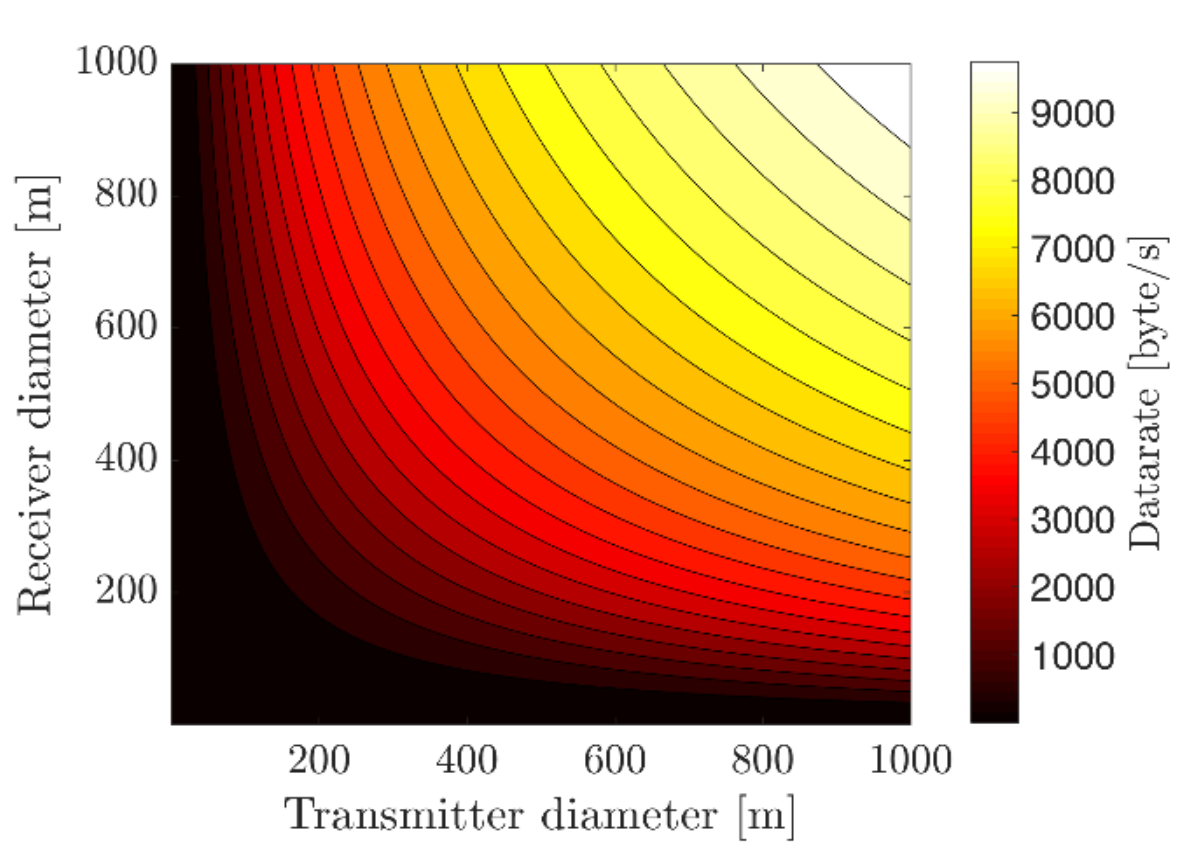

Figure 3: RF communication design space at 4.3 lightyears distance using the same assumptions stated in the text above with 1W transmitter power.

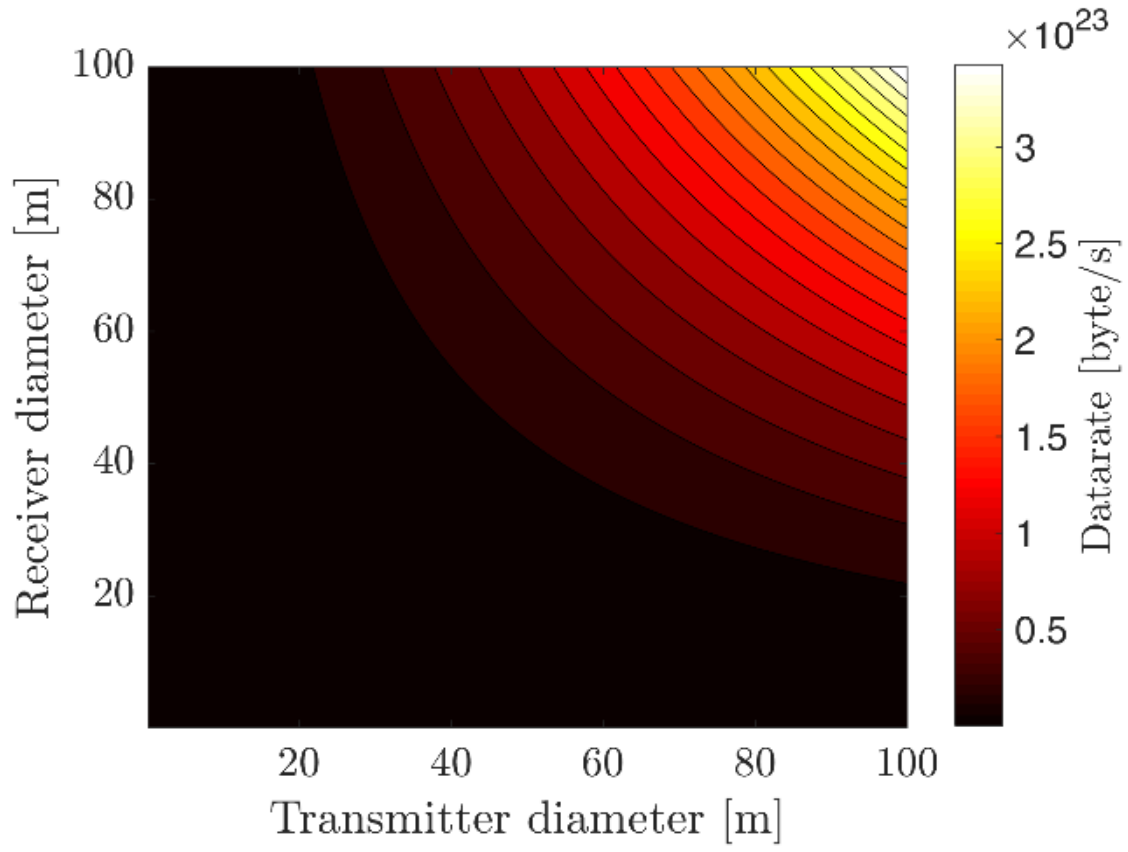

Figure 4: Optical communication design space at 4.3 lightyears distance using the same assumptions stated in the text above with 1W transmitter power and 1 bit=1 photon.

An option to maximize gain on the transmitter end and to avoid additional mass/spacecraft components, is to utilize the laser sail itself as the antenna, for both RF and optical communication.

A completely different, unconventional communication method is to directly use environmental features for communication. Like exoplanet search, where the occultation of the stellar disk is used for detecting planets via the transit method, we can imagine that a probe flying towards Alpha Centauri occults one of the two bright stars. By using electro-chromic patches on the laser sail, the sail can change its optical transparency. Given the natural variation in luminosity of the star, we can imagine artificial luminosity changes via the sail that are strong enough to be detected from the Solar system. Such a communication system would transmit a form of Morse code and would have a very low data rate. However, it might be possible to transmit basic telemetry data.

Using trailing spacecraft as relay stations for communication the distance for communication would be reduced, however, the mission complexity overall increases due to the sequential launch of the spacecraft.

As this section shows, there are several options for communication over interstellar distances that are in principle feasible with technologies likely to be available during the next 10-20 years. Moreover, there is potential for several augmentations that can be made to classical communication system to make such systems work at the gargantuan scales we are considering.

## 2.3 Spacecraft Probe Size (Miniaturization)

The sorts of electronics that can be packed into a small probe was analysed. Our starting point was ChipSat (sprite) technology [11], [35].

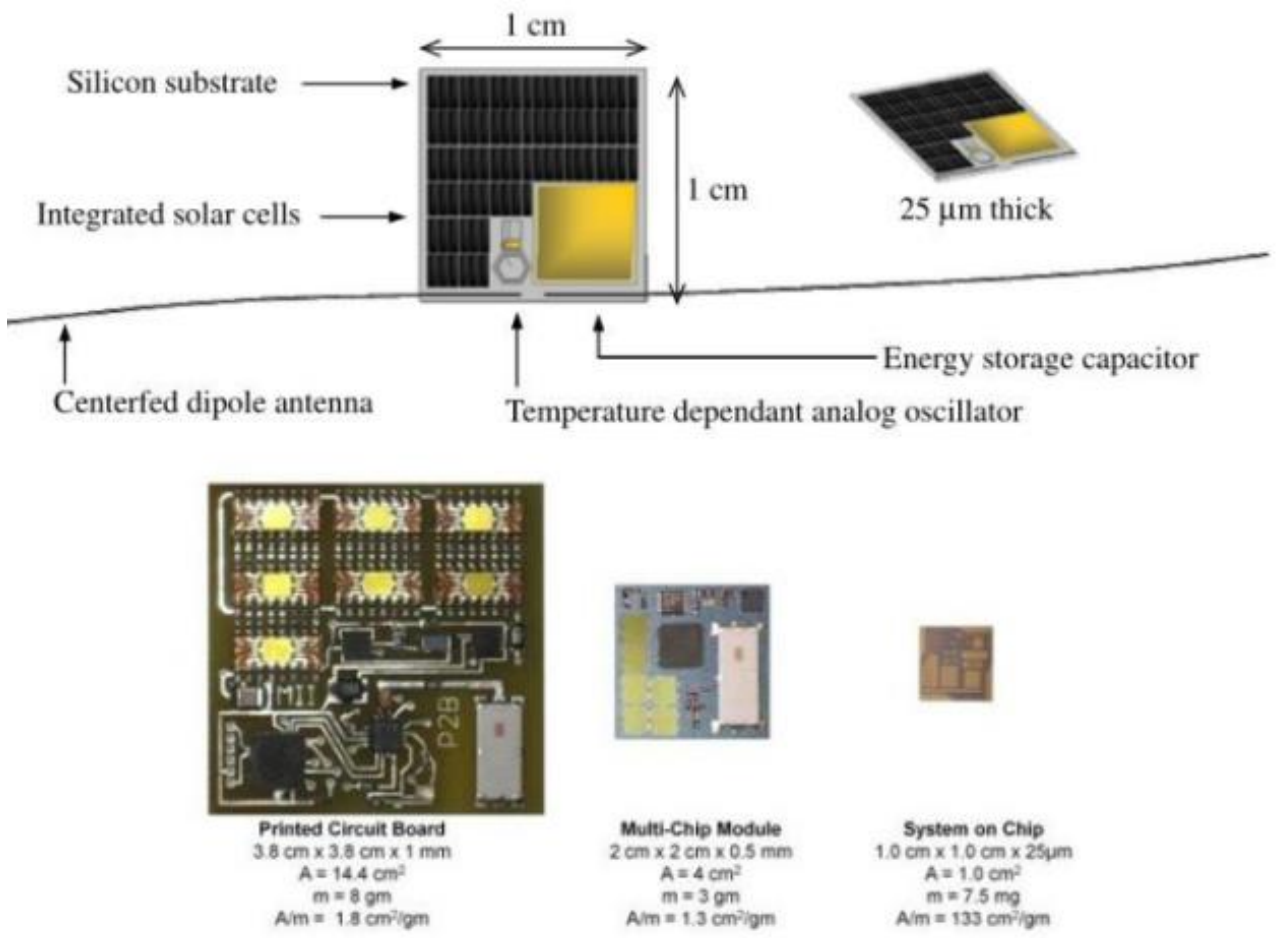

*Figure 5: Chipsat Probes*

Future trends in miniaturization may enable unprecedented spacecraft, mission architectures, and laser infrastructures:

- ***Swarm lenses***, created by spacecraft swarms such as proposed in the NASA NIAC "Orbiting Rainbow" concept, as introduced in Section 2.18.1. Such a concept could allow for launching a large number of "swarm" spacecraft in the gram-range and still send data back. Required laser power could be significantly reduced that way. Furthermore, massive redundancy could be established [36]. Another approach was proposed by Underwood et al. [37] by assembling a large optical telescope from small autonomous telescope elements. Each element is a CubeSat, equipped with the basic satellite subsystems.
- **3*D-integrated electronics*:** Circuits and sensors could be further integrated via 3D-integration, further reducing the mass of the spacecraft [38], [39].
- ***Miniaturization:*** Further miniaturization in electronics and the use of NEMS (nano) components could allow for even smaller spacecraft sizes [40], [41]. Fundamental physical limits for optical instruments still exist [42], however, such instruments could be formed by distributed small spacecraft, such as chip-sized satellites forming a lens and another spacecraft carrying a detector [43]–[46]. Intuitively one would expect that the miniaturization of electronic components would increase its vulnerability to the interstellar radiation environment, as the damage from the impact of a single radiation particle would affect a higher number of components. However, recent advances in Field-Programmable Gate Arrays (FPGAs) and latch up protection have considerably reduced the vulnerability of electric circuits to in-space radiation in general.

## 2.4 Stability of Sail to ride the Beam

Passive sail stability can be achieved by a conical shape of the sail, as [47]–[49] have shown analytically and experimentally for microwave sails. The resulting force vector moves the sail automatically to the centre of the laser beam. However, such a sail geometry does not seem to be appropriate for sails that are subject to high acceleration loads. Hence, more recent results [50] have proposed a spherical sail geometry. As the authors have remarked, such a geometry would increase the cross section of the spacecraft in flight direction, increasing the amount of interstellar matter impacting the sail and the risk of a catastrophic dust particle impact. Further open questions are which geometries are optimal for laser sails, depending on the laser and sail material characteristics, as [5] has demonstrated for different laser array geometries.

## 2.5 Thrust Vector Control of Sail

Several attitude determination and control system (ADCS) options exist during and after the acceleration phase. Attitude control refers to the change in attitude of the spacecraft. Determining the attitude of the spacecraft is often a necessary condition for attitude control. By contrast, changes in position are referred to as "trajectory modifications".

Attitude control during acceleration can be accomplished via several technologies: Electro-chromic surfaces such as used for the Ikaros probe is one option, displacing a mass on a boom, and small flaps at the outer edges of the sail are alternatives [51]–[53]. Electro-chromic surfaces change their optical properties by changing electric potential [54]. A darker part of the sail reflects less and absorbs more; hence a momentum is generated if the symmetric opposite part of the sail is not darkened. Displacing a mass on the boom results in generating a momentum, as the centre of mass of the sail changes the direction of the resulting force vector on the sail [52]. Small steerable flaps on the edges of the sail create small forces on the sail that induce torque and a momentum [51].

***Options during acceleration phase:***

- ***Electro-chromic surfaces on sail*** [54], [55]: The area with the electro-chromic coating changes its optical property. Usually the electro-chromic surface decreases its reflectivity. This means that the resulting force acting on the side with the coating is lower than on the opposite side of the sail. The resulting torque turns the sail around the x-axis.
- ***Mass on boom*** [52]: Another option for inducing a torque is to displace the centre of mass of the sail. The change in centre of mass results in a larger angular momentum on the left side of the sail in the illustration than on the right. In consequence, a torque is generated that turns the spacecraft around the axis perpendicular to the plane of this document.
- ***Flaps*** [51]: Using small flaps at the edges of the solar sail generates a torque on the sail by changing the angle of the sail. Contrary to the two former options the flaps enable to induce a rotatory movement to the sail by creating a torque around the axis going through the centre of the sail, similar to a windmill.

***Options after acceleration phase***

Controlling the spacecraft after the acceleration phase is challenging but still possible. A classic approach is to use thrusters with a high specific impulse in order to save fuel and at the same time a low mass. One option is the field-emission electric propulsion (FEEP) that is already being developed for CubeSat's [56], [57]. FEEP specific impulses range from 4,000s to 8,000s and thrusts in the µN range. Propulsion-less systems are, for example, electromagnetic tethers and electric sails [58], [59]. By deploying tethers or wires asymmetrically, a torque can be induced via the momentum of incoming interstellar electrons and protons. A further option is the use of the galactic magnetic field, which was rather proposed for trajectory modifications in the past [15], [16]. Due to the low field strength of the magnetic field of an average of 1nT compared to 30µT for the Earth magnetic field [60]–[62]. Assuming a wire with a diameter of 1m and 1 to 1000 windings and a 0.1A current, an angular momentum on the order of $10^{-10}$ to $10^{-7}$ Nm can be generated. The induced angular momentum would be one to several orders of magnitudes lower than for the FEEP thruster. As another option, [5] proposes photon propulsion. It would create thrusts in the nN range.

| **Technology** | **Propellant needed?** | **Thrust range** |
|---|---|---|
| *FEEP* | Yes | $10^{-6} – 10^{-5}$ |
| *Electromagnetic tether* | No | Depends on current and length |
| *Electric sail* | No | Depends on current and length |
| *Magnetorquer* | No | $10^{-7} – 10^{-10}$ |
| *Photon thrusters* | No | $10^{-9}$ |

## 2.6 Spacecraft Instrumentation and Sensors

Recent advances in MEMS technology have resulted in a plethora of miniaturized sensors for space applications [63]–[66]:

- ***Printed electronics*** [27]: There are a range of sensors such as temperature sensors that can be printed via technologies commonly used in the area of printable electronics. These electronics are usually limited by their complexity and voltage. Whereas standard space electronics operates at 3 V or 5 V, printed electronics would operate at higher voltages such as 30 V. Furthermore, due to limitations in printing resolution, more complex components such as microprocessors cannot be fabricated with this technology. However, the main advantage of printed electronics is its low weight. In a NASA NIAC study [27], a basic miniature planetary probe prototype was designed with a total mass of 4 grams.
- ***Micro Electric Mechanical Systems (MEMS) sensors:*** A large variety of MEMS sensors exists such as:
  - Magnetometers [67], [68]
  - Gyroscopes [69]–[71]
  - Accelerometers
  - Sun sensors [72]
  - Star tracker [23], [73], [74]
  - Sun sensor [72]

It is clear that there are basic physical limits to miniaturization for certain instruments such as the aperture size for optical instruments. An overview of possibilities for miniaturization for CubeSat applications is given in Selva et al [42].

For our spacecraft, we propose the following set of instruments:

- *MEMS spectrom*eter [75]–[77]
- *Optical MEMS camera with aperture:* Used as star tracker and camera. The aperture is made of a graphene sheet structure that is extremely light but is still a good stiffness characteristic.
- *Camera:* standard CMOS technology
- *MEMS particle detector*
- *MEMS magnetometer*

***Miniature Gamma Ray Spectrometer***

It is proposed to use a Sodium Iodide (Thallium doped) scintillating crystal, wrapped for protection with PTFE tape on 5 faces, and having the 6th face bonded to a large area photodiode. This is then connected to a sensitive amplifier. High energy impacts with the crystal will result in scintillation, and this scintillation will be seen by the photodiode. A spectrum will be produced. This is a known method for use in small space spectrometers.

***Optical Camera (doubling up as star tracker)***

A smartphone camera module will be used to reduce mass, since these modules have masses as little as a few grams. Using a camera sensor with the largest Megapixel count, however, would not be the most effective solution. What is required is a camera module which has less noise than other modules, whilst still producing an adequate resolution image. It is anticipated that a camera module of about 8 to 12 Megapixels will be chosen for this reason, and because of the mass constraints, a lens system based on the lenses used as external lenses for smartphones would be optimised for this purpose.

***IR Camera***

The IR camera would consist of a smartphone camera module, similarly to the optical camera. It would only include the necessary filter for use in the near IR.

***Light Intensity sensor***
A Light Intensity sensor measures the intensity or brightness of light. In this application, it can be used as a way to determine the brightness of a target object being pointed at. A Light Intensity sensor capable of detecting as little as 188 uLux up to 88,000 Lux, and with a 600,000,000:1 dynamic range has been identified. This sensor has a mass of 1.1g

***UV sensor***
The UV sensor proves another emission detecting sensor which would be pointed at the appropriate target with the other sensors. The baseline UV sensor is a UV-B Sensor with a 240nm - 370 nm range, and with a mass of 0.7g.

***Temperature sensor***
There are numerous surface mount temperature sensor solutions, and it is proposed to use one of these temperature sensors, with a suitable operating range.

***Radiometer***
It may be possible to incorporate a compact radiometer, however work is ongoing to determine the packaging constraints for this sensor.

***Inertial Navigation***
The Inertial Navigation System consists of:
3solid state gyros (X, Y and Z-Axis)
3 solid state accelerometers (X, Y and Z-Axis)
3 solid state magnetometers (X, Y and Z-Axis)
Star Tracker (the optical camera)
Sun sensor (used for tracking the target star during fly-by)

The well-known problem of drift can be mitigated by using the star tracker and a Kalman filter, along with the temperature sensor. These will enable the drift of the gyros to be zeroed out.

***Sensor Positioning***
All imaging or related sensors would be orientated in the same direction, to ensure they are all imaging the same target. This would include the optical camera, IR camera, light intensity sensor and UV sensor.

## 2.7 Interstellar Dust and High Energy Particle Bombardment

The probe is subject to the interstellar medium it traverses during its flight. The effects of the interstellar medium on the spacecraft have been previously explored by [78] and [79]. Three types of effects of the interstellar medium on the spacecraft can be distinguished: Galactic cosmic rays, particles, and dust [80]. The galactic magnetic field has already been previously introduced.

The first is high-energy cosmic rays which derive from either solar energetic particles or galactic cosmic rays. Of this, 85% are high-energy protons, 14% are alpha particles, and the remainder is small portions of positrons and antiprotons (<1%). The peak of this energy spectrum tends to be at around 0.3 GeV. High-energy cosmic rays can alter the states of electronic circuits. The mitigating options to manage this include physical shielding (e.g. radiation hardening), magnetic shielding or some combination of both. A spacecraft in interstellar space is subject to a higher flow of heavy particles as inside the Heliopause [81], [82]. On the other hand, Solar protons are absent. For an interstellar probe, it is essentially impossible to protect against galactic cosmic rays, except by more than a meter of hydrogen or lead.

Interstellar particles predominantly consists of hydrogen and electrons. Hydrogen occurs in either a neutral or ionized state. The protons can have significant affects at velocities greater than 0.01c, where energy is dissipated into the structure through atomic ionisation leading to emission of radiation and structural heating through absorption. At very high energies these collisions may cause plastic deformation and local melting, leading to a degradation of the spacecraft structure. The electrons from the interstellar space will lose their energy by emission of radiation as an electron passes close to the nucleus, and radiation is absorbed by the material structure via the photoelectric effect, Compton scattering and pair formation. The impact of the electrons may also lead to the emission of Bremsstrahlung radiation as soft x-rays, which further heat the structure. Estimates for the interstellar hydrogen, proton, and electron density have been presented in [80] and are depicted in Table 0-1.

*Table 0-1: Interstellar hydrogen and electron density ranges* [80]

| **Interstellar medium element** | **Conservative density range [particle/cm³]** |
|---|---|
| Atomic hydrogen | 0.10 – 0.23 |
| Electrons | 0.05 – 0.21 |
| Neutral and ionized hydrogen | 0.15 – 0.44 |

Interstellar dust grains are substantially smaller than interplanetary dust grains. A typical grain size is 0.1μm. Interstellar dust comprises about 1% of the mass of the interstellar medium, where 99% are hydrogen and electrons. However, the energy of an impacting dust grain on the structure is much higher than for protons or electrons. When the particle slows down into the material to subsonic speeds, the kinetic energy is deposited into a volume larger than the particle itself. Hence the energy density and its temperature are much higher, around $10^{12}$ K. This is high enough to vaporise and cause mass loss through ablation. For a typical grain size $10^{-16}$ kg the impact energy of order $10^{11} – 10^{12}$ MeV at 0.15c. Two processes will then result from such bombardment, which includes heating of the material but no permanent damage, and processes which will cause permanent damage.

As examples to study we considered three types of geometries, which we call model 1, 2 and 3 and all are assumed to be cylinders. Model 1 has a radii of 1 mm and a length of 10 mm (aspect ratio 2: 10) which we call the 'small model'. Model 2 has a radii of 10 mm and a length of 100 mm (aspect ratio 20: 100) which we call the 'slim model'. Model 3 has a radii of 20 mm and a length of 100 mm (aspect ratio 40:100) which we call the 'fat model'.

One of the considerations is simply the temperature rise due to the incoming protons and electrons, but permanent changes in the material will only occur if the temperature is sufficiently high. For the three geometries considered we find that the temperature is in fact not high and amounts to 192.7 K for the 1 mm radii design and 108.38 K for the 10 mm and 20 mm radii designs. The surface temperature is mainly a function of the frontal area geometry and the incoming energy flux.

It is necessary to assess the mass loss and therefore frontage shielding requirements on our mini-spacecraft. To do this we can adopt the Benedikt [83] relation or the Langton [84] relation and we will use to make an approximate assessment for how much shielding is required. We will look at three different areal geometries for our 280 gram probe. We will also look at three different temperatures for comparison, which includes 600 K, 1,000 K and 1,500 K.

The Benedict relation is an equation to model the mass loss due to material erosion from an incoming particle stream. It is defined by several parameters including the fraction of the energy that is transferred from the medium, which will result in permanent changes in the material of the vehicles – this is assumed to be 0.25. It is also derived from the heat flow required to destroy a unit mass of the material. The latent heat of sublimation defines the characteristic property of the chosen material. The equation is also a function of the velocity relative to the speed of light. The frontal area of the probe also features in this relation.

Some calculations were conducted to assess the mass loss from the probe under the assumption of a 0.1c velocity for a 50 year mission profile. A crucial factor in these calculations is the assumption of interstellar medium density. Note that the conservative upper and lower bounds for the density of hydrogen and electrons is $2.49 – 7.30*10^{-22}$ $kg/m^3$. The results of these calculations for the three model geometries are shown in the table below and these are conducted assuming a Beryllium frontal shield material.

***Table 0-2: Particle Shield Calculations for a Beryllium material over a 50 year flight at 0.1c.***

| Model | $2.49*10^{-22}$ kg/m³ | $7.3*10^{-22}$ kg/m³ |
|---|---|---|
| 1 | $dm/dt=3.125\times10^{-12}$ kg/s, 5 g mass ablated | $dm/dt=9.162\times10^{-12}$ kg/s, 14.6 g mass ablated |
| 2 | $dm/dt=3.125\times10^{-10}$kg/s, 493 g mass ablated | $dm/dt=9.162\times10^{-10}$ kg/s, 1445 g mass ablated |
| 3 | $dm/dt=3.125\times10^{-10}$ kg/s, 1972 g mass ablated | $dm/dt=9.162\times10^{-10}$ kg/s, 5781 g mass ablated |

***Table 0-3: Particle Shield Calculations for a Graphite material over a 50 year flight at 0.1c.***

| Model | $2.49*10^{-22}$ kg/m³ | $7.3*10^{-22}$ kg/m³ |
|---|---|---|
| 1 | $dm/dt=1.850\times10^{-12}$ kg/s, 2.95 g mass ablated | $dm/dt=5.425\times10^{-12}$ kg/s, 8.65 g mass ablated |
| 2 | $dm/dt=1.850\times10^{-10}$kg/s, 292 g mass ablated | $dm/dt=5.425\times10^{-10}$ kg/s, 856 g mass ablated |
| 3 | $dm/dt=1.850\times10^{-10}$ kg/s, 1168 g mass ablated | $dm/dt=5.425\times10^{-10}$ kg/s, 3423 g mass ablated |

Overall these numbers look very encouraging in terms of our aim and for the reference design we recommend studies start with the assumption of a 20 grams shield provided the length of the probe is much greater (×10) than the radii on the frontal area. For comparison it is worth noting the heat of sublimation rates of some standard materials that are potential shield candidates. This includes Lithium ($2.57\times10^6$ J kg/m), Beryllium ($35.53\times10^6$ J kg/m), Boron ($53.6\times10^6$ J kg/m), Graphite ($60\times10^6$ J kg/m), Aluminium ($12.1\times10^6$ J kg/m). Some examination of the material aerogel is also recommended which could have ideal applications as a heat shield given its lightweight structure.

One important assumption of the above analysis is n, the fraction of energy converted from impacts that does permanent damage. We assumed a high value of 0.25 but this was very conservative as a worst-case model so there does appear to be plenty of scope for reducing the shield mass overall.

It is assumed that the distribution of interstellar dust size follows the power law distribution [85]. Hence, there is a chance that the probe might collide with a dust particle that would instantly destroy it. This risk can be mitigated by redundant probes. In that it is not one probe that is sent but many, and this way if any are lost it would not matter.

So with our small probe one of the problems is, how to get smaller and still protect sufficiently against lattice displacement. We recommend the adoption of an ablative material on the front of the probe. This should be a material that is extremely light but would still absorb the impacts. The key requirement for any material is that they are very lightweight but also subject to mass ablation so they can be eroded away by the constant stream of particles.

However, this ablative front shield will only mitigate the incoming high-energy particle impacts. What about the radiation protection given that any electronics will likely have a krad radiation limit beyond which they will not work? For this it may be necessary to explore the option of generating a self-induced magnetic field around the probe, as a form of mini-magnetosphere. But this will not mitigate the most energetic of particles.

Also this comes at the cost of increasing the mass to the vehicle. Again, playing the statistics game by having many probes as a form of redundancy should take care of that and the objective would be to get a large percentage of the probes to complete the journey.
In addition, the on board computer should be coded up with fault-tolerant software, which would decouple damage done to the hardware from leading to software errors. Third, we advise that the electronics is switched off most of the time so that any flip switching will not have any effect on the probe.

There are developments today in relation to CubeSats that will be immune to a lot of radiation. It is possible to implement any processor on an FPGA, with the benefit of being able to re-route your FPGA in space and therefore “reset & heal” your own main processor. If you then mix that with some radiation hardened memory (maybe MRAM, maybe phase change memory, maybe FRAM or something else) and you can run your system for a long time, because you are able to re-route your processor on the FPGA. There are small FPGAs that have already been developed and these may have enough processing power to run a really low level program for us. And maybe we can use some radiation hardening manufacturing process on them (against SEL, SEB, and all the hard errors). We can also implement a lot of strategies to cope with radiation effects through software. Especially the MRAM with the critical information can be secured. Overall we recommend the implementation of a hybrid protection system that includes radiation hardening, self-healing capability and corrective software. As this will add mass to the system so trade-studies will need to be constructed.

## 2.8 System Reliability over Mission Time

An obvious approach to reliability and redundancy is simply to propel multiple (thousands or millions) of probes towards the nearest stars. This also has the advantage of a potential long distance space communications option, in terms of utilising optoelectronics to encode information on the return wave of a reflected beam. The array, would present an effective reflection disc.

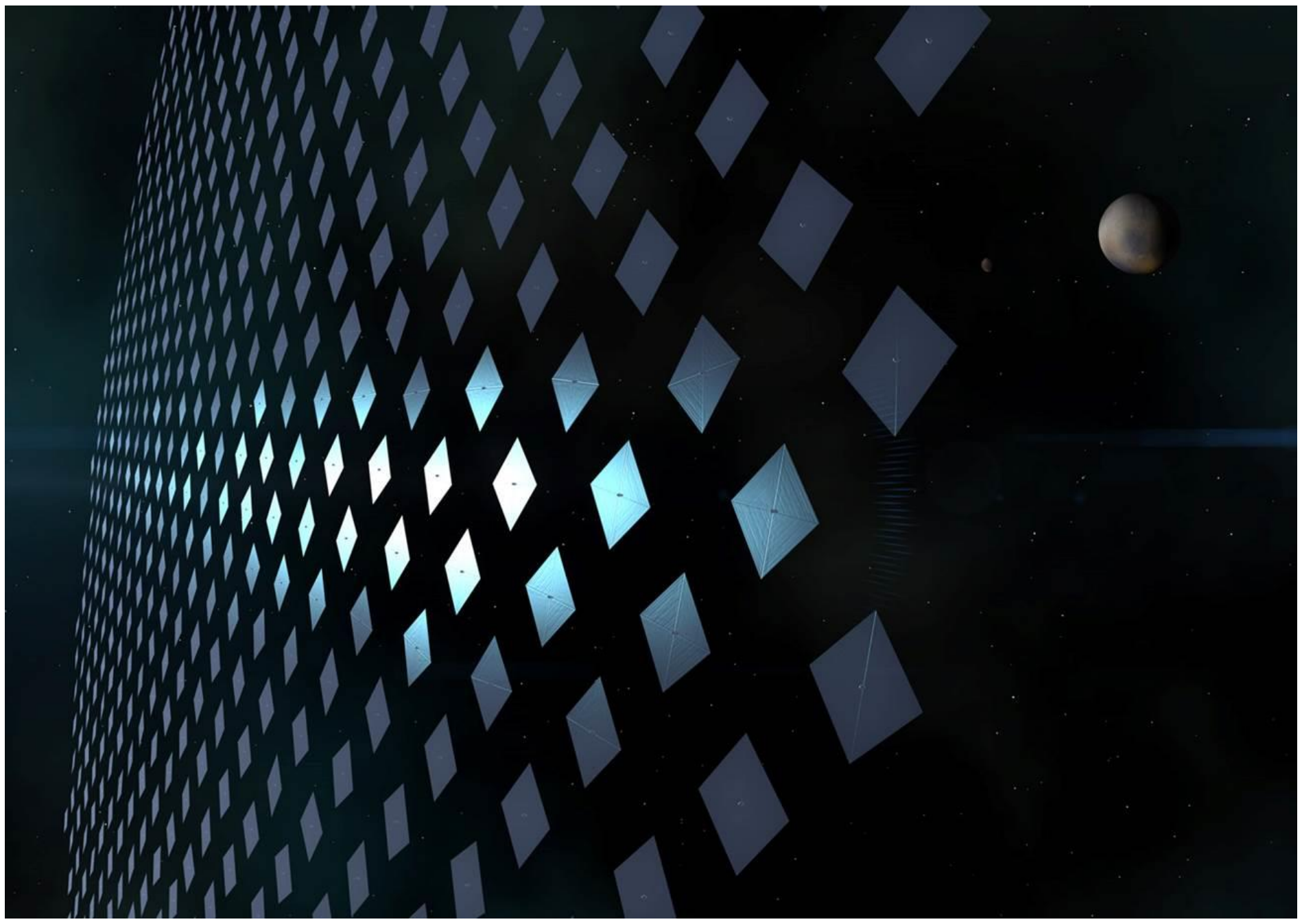

*Figure 6: Illustration of multi-sail array for reliability, redundancy and enhanced communications link (Credit: Adrian Mann)*

We assume that the spacecraft remains in hibernation mode for most of its trip. Due to the power limitations imposed by the nuclear battery and the supercapacitor, subsystems are activated selectively. Note that hibernation protects from life-time issues from operation but does not prevent effects that are independent of the component operating. For example, radiation degradation takes place irrespective of component operation.

Apart from hibernation, the following mitigation strategies are used, as already introduced in an earlier section:

- Minimize mechanical components that are subject to wear: magnetically levitated MEMS gyros
- Redundancy: Although we assume that one spacecraft is a single-string system, we assume the launch of several spacecraft of the same type.
- FPGA: For protecting against software errors induced by radiation, we rely on FPGAs, which are very tolerant to radiation effects on the hardware. The software can be executed even with damages inflicted to the hardware.

## 2.9 Spacecraft Thermal Issues

Use of radioisotope battery (power subsystem) to keep spacecraft heated. The nuclear battery generates 1 We with an efficiency of 5% [86]. The remaining 95% of the energy generated is released as heat. This is about 19 W of heat which is transmitted via heat conductors such as copper circuits to components that need to be kept above certain minimum temperatures. As a more recent technology, carbon nanotubes could be used for transporting heat [87], [88]. Whenever components need to be operated and woken up from hibernation, the heat from the battery could be selectively transmitted to these components allowing for a rapid wake-up from hibernation. The On Board Data Handling (OBDH) system is located close to the battery in order to allow for a radiative heat transfer independent of heat conducting circuits.

## 2.10 Power Supply for the Probe

In consideration of the power supply for the probe we have identified several options which will need further examination. These options can be divided into options relying on power sources external and internal to the spacecraft.

Internal power sources are devices which are brought on board of the spacecraft and produce energy without any external inputs. One of the technologies suitable for the interstellar probe could be a **scaled down RTG** utilising Americium-241. This element has a radioactive half-life of 432 years, which is longer than the one of Plutonium 238 (87.7 years), but comes at a cost of a reduced power density of 0.13 kW/kg, which is low compared to the standard Pu-238 having a power density of 0.54 kW/kg. However, Americium-241 also produces more penetrating radiation through decay products so increased shielding may be required. There are prototype RTG designs that can be examined which give output of 2-2.2 We/kg for 5 We RTG designs.

Landis et al. [89] have investigated the use of **alphavoltaics** in space. Contrary to RTGs that are based on thermoelectric conversion, alphavoltaics convert the ionization trail of alpha particles into electricity. Commercially available alphavoltaic cells have a specific power of about 0.33W/kg [89] which is three orders of magnitude less than for RTGs. Landis et al. [89] have also recently investigated the use of **betavoltaic** cells for small space missions. Betavoltaics use a direct conversion approach by exploiting elecone-hole pairs that are generated by electrons emitted from the radioactive material. The main shortcoming of betavoltaics is the use of low power density and the short half-life of the used radioactive substances such as Tritium and Promethium-147. Furthermore, the use of nuclear D-cell battery is an option. A 5-20 We range is being developed by the Centre for Space Nuclear Research [90].

In contrast to alphavoltaics and betavoltaics which directly make use of the emitted radiation of a radioactive source, **radioisotope thermophotovoltaic cells** convert the radiation from a radioactive material via photovoltaic cells in the infrared spectrum [91]. The main advantage is the higher efficiency, compared to other nuclear batteries. Teofilo et al. [92] report an integrated specific power of 14We/kg for thermophotovoltaic cells with efficiencies between 19-25%. Hence their specific power is about two orders of magnitude higher than for RTGs.

Another example worth mentioning is the possibility of developing a tiny battery powered by **micro-organisms.** Microbial fuel cells will rely on some kind of organic matter and would have limited lifespan – and currently we do not know how long microbes can metabolise in extreme environments. Yet, they can generate small amounts of electricity. Such a system could be biofilm powered [93], [94]. Recently developed miniature microbial fuel cells generated electric power on the order of $10^{-6}$W at a volume of 0.0026mm². Using the density of Carbon (the fibers used are made of Carbon) yields a specific power of 788W/kg. The specific power is therefore about the same order of magnitude than for radioisotopic thermal generators. However, some drawbacks are the expected lower stability of the power source over long durations, and the need for consumable mass.

A comparison of the aforementioned technologies is presented in Table 0-4. For this feasibility study, the Strawman requirements are used, leading to the assumption of 0.05% transfer time and 99.95% charging time for the power system. With a 5W transmitter power and an efficiency of 50%, the required input power is 10W and this yields a 0.02 W continuous power need. The nuclear D-cell battery is identified as the most promising technology with higher feasibility due to the low mass.

*Table 0-4: Comparison of internal power sources.*

| **Technology** | **Feasible ?** | **Power density** |
|---|---|---|
| RTG | No (too heavy: ~0.01 kg) | 2-2.2 We/kg |
| Alphavoltaics | No (too heavy: ~0.06 kg) | 0.33 We/kg |
| Betavoltaics | No (too heavy, too short half-life of Tritium, Promethium 147) | - |
| Microbal battery | No (stability, temperature) | - |
| CubeSat Nuclear D-cell battery (thermophotovoltaics) | Yes (~0.0013-0.0017 kg) | 12-16 e/kg |

Options with power sources external to the spacecraft can either rely on artificial or natural sources. Artificial sources imply beaming electromagnetic waves from the solar system to the sail as a remote power source, as proposed by Forward [1]. Either a **laser beam** or a **microwave beam** could be utilised. In the case of beamed-laser power, getting 0.1 W/m$^2$ to the spacecraft over a distance of 4 light years would require 10 GW of laser power when assuming a wavelength of 1000 nm and nrad pointing accuracy. For a 1000km spot size, a 90km aperture at the solar system would be needed. For the beamed microwave-power (Starwisp), very high power levels and receiver antenna are needed. Due to the large wavelength, the size of the aperture is also prohibitive making the option rather unattractive.

Various natural sources are also available in the interstellar space. Stellar luminosity, i.e. residual light from nearby stars could be captured by solar cells. The areal power density of these light sources is likely on the order of $10^{-7}$ to $10^{-6}$W/m², if it is assumed that 5 stars illuminate the spacecraft from a distance of 100,000AU. Use of an ultrathin and lightweight organic solar cell could be made for capturing the rays. Current research [37] claims 0.1 kg/kW so that for a 100 W system you would only need 10 g of solar cells when close to a star. The resulting captured energy of the starlight is rather low and it would take months for harvesting 1 Ws with a 1m² surface area. The large diffusion of this light makes inefficient for powering the probe.

Fluctuations in the **galactic magnetic field** could also be exploited: Due to the turbulent and anisotropic structure of the galactic magnetic field, changes in the field strength could be exploited for generating a current via the Lorentz force. The field strength is close to 0.1-0.6 nT. However, the gradient of the field strength does not seem to be known today and it is difficult to estimate how much power could be generated from a system exploiting it. **Galactic cosmic rays**, protons, and electrons could be used for creating an electric current. This could be done by transforming the kinetic energy of the incoming particles and rays into electric energy with the help of betavoltaics. Assuming that the kinetic energy of incoming charged particles could be completely converted into electric energy, at 10%c, a surface area of 1cm², a power between $10^{-15}$ and $10^{-16}$ W could be harvested. In reality, the power is likely orders of magnitude below these levels. Hence, using interstellar particles as an energy source does not seem to be a viable option, unless a significantly larger area is covered.

Finally the use of an electromagnetic tether as proposed by Matloff is quite promising. We can supply some 5 W during flight with a 100 m tether by capturing interstellar electrons. The tether can be broken up into 10 x 10 m tethers. The current generated is about 3.3 Amp. The mass of the tether is below 0.1 gram. The whole system could be built for a mass close to 0.1 gram [59].

A summary of the technologies is shown in Table 0-5, where it is evident that the electromagnetic tethers and the beamed laser power are the most mature options.

*Table 0-5: Comparison of external power sources.*

| **Technology** | **Feasible ?** | **Power density** |
|---|---|---|
| Solar cells (Stellar light) | No (yes close to star) | $10^{-6}$ – $10^{-7}$W/m² @interstellar space |
| Galactic magnetic field | No | Unknown fluctuations |
| Cosmic rays with beta voltaics | No | 10-11-10-12W/m² @0.1c |
| Electromagnetic tether (Charged interstellar particles) | Yes | 5W @10x10m tethers; 0.1gram |
| Beamed laser power | Yes | 0.001 W/m² or higher, depending on spot size |
| Beamed microwave power | No | |

Clearly a small research program will be required to work out the best system for the interstellar probe but what is encouraging is that there are lots of options that already exist or are emerging to supply the on board power. This issue does not appear to be a show stopper.

## 2.11 Data and Telemetry Management

Data from the sensors is temporarily stored in the memory within the OBDH system. Once the communication downlink to the Solar System is established, the stored data is then forwarded to the communication system. Currently, no on board-treatment of the data is planned, as this would probably exceed the hardware and software capabilities. However, with future advances in computing and electronics, data treatment on-board could be envisaged in order to reduce the amount of data transmitted by selecting the most relevant data.

## 2.12 Sail Material and Design

The laser sail material is a crucial element for the success of a small interstellar probe as its temperature limit determines the maximum laser power per surface area and its areal density determines the mass of the sail. The maximum temperature depends on the optical characteristics of the material. Hence, the areal density and the optical characteristics are the most important parameters. The values from Matloff, 2012 are shown in Table 0-6 [7]. The analysis conducted by the team from the Technical University of Munich [95] during the Project Dragonfly design competition resulted in the choice of a graphene sandwich sail due to its superior thermal characteristics for resisting the laser beam and its extremely low density. One disadvantage of graphene-based sail materials is the low reflectivity compared to materials such as Aluminized Mylar. However, the extremely low density compensates for this shortcoming.

*Table 0-6: Comparison of Aluminized Mylar, Grahene Monolayer and Graphene Sandwich according to Matloff, 2012* [7]

| Parameter | Aluminized Mylar (Edwards et al. 2002) | Graphene Sandwich | Graphene Monolayers doped with Alkali Atoms |
|---|---|---|---|
| **Reflectivity** | 0.9 | 0.05 | 0.05 |
| **Absorptivity** | 0.1 | 0.03 | 0.4 |
| **Emissivity** | 0.075 | 0.03 | 0.03 |
| **Operating Temperature** | 500 K | 4500 K | 450 K |
| **Thickness** | $2x10^{-6}$ m (with 50 nm Al film) | $0.335x10^{-9}$ m | $0.5x10^{-9}$ m |
| **Areal Density** | $2.7x10^{-3}$ kg/m² | $7.4x10^{-7}$ kg/m² | $5x10^{-6}$ kg/m² |
| **Calculated Power Density** | $5.3145x10^{3}$ W/m² | $4.6501x10^{7}$ W/m² | $2.79x10^{2}$ W/m² |

A technology which can significantly improve the performance characteristics of an interstellar sail is the one of dielectric materials. The main benefit of dielectric materials lies in their ability to have their reflecting properties "tuned" at a specific wavelength. By alternating between high index and low index dielectrics, the reflectance at a specific wavelength can be increased close to unity (Forward 1986). Their high emissivity, low absorption and high-temperature properties makes them suitable for high laser intensities. (Landis 1999)

The reflectivity of a dielectric material reaches a maximum when the thickness of the film $t$ is equal to one quarter of the light's wavelength within the film $(\lambda/n)$. This is when constructive interference can occur and the condition reads:

$$t = \frac{\lambda}{4n}$$

The reflectivity $R$ of the quarter-wave layer film is then given by (Landis 1999):

$$R = \left(\frac{n^2 - 1}{n^2 + 1}\right)^2$$

A summary of promising dielectric materials based on oxides is given in Table 0-7.

*Table 0-7: Dielectric sail material properties*

| Material | Reflectivity [-] | Density [kg/m³] | Emissivity [-] | Max. Temperature [K] | Max intensity [W/m²] |
|---|---|---|---|---|---|
| *Al2O3* | 0.26 | 3960 | 0.9 | 2500 | 2.7 10^6 |
| *Ta2O5* | 0.52 | 8750 | 0.25 | 2140 | 6.2 10^5 |
| *ZrO2* | 0.42 | 5500 | 0.95 | 3000 | 7.5 10^6 |

The values for reflectivity, density and maximal temperature are taken from Landis 1999. With the values found in literature for the emissivity, the maximal intensity was calculated.

Further increasing the number of layers of the dielectric film can lead to even higher reflectivity, which leads to two effects: the force on the sail increases and the absorbed energy becomes lower, leading to less strict requirements for the thermal management. At the same time however, the total mass of the sail is increased leading to an optimum for the number of layers.

Lubin has proposed the superposition of multi layer dielectrics on metalized plastic film, leading to reflectivity close to 99.995 %. The dielectric layers are comprised of $SiO_2$ and $TiO_2$, whereas for the metalized film Ag and Cu have been tested.  For multi layer dielectric on metalized glass reflectivity values up to 99.999% have been achieved. These values refer to a short bandwidth around 1.06 microns of laser wavelength.

A summary of   the tested materials and their respective properties is given in Table 0-8.

*Table 0-8: Dielectric multi-layer sail material properties*

| **Material** | **Reflectivity [-]** | **Density [kg/m$^3$]** | **Emissivity [-]** | **Max. Temperature [K]** | **Max intensity [W/m$^2$]** |
|---|---|---|---|---|---|
| *Ag, SiO2, TiO2 5 layers* | 0.9961 | 1400 | 0.03 | 1235 | 1.0 10^6 |
| *Ag, SiO2, TiO2 15 layers* | 0.9999535 | 1400 | 0.03 | 1235 | 8.5 10^7 |
| *Cu, SiO2, TiO2 15 layers* | 0.9999294 | 1400 | 0.07 | 1360 | 1.9 10^8 |

The sail has a slightly conical shape in order to make it self-stabilizing when propelled by the laser beam. If the sail is subject to perturbations and is displaced from the laser maximum, it would automatically create a thrust force to realign with the beam maximum. This principle has already been introduced in an earlier section. Table 0-9 shows the main parameter values of the sail.

***Table 0-9: Laser sail characteristics***

| **Characteristic** | **Value** |
|---|---|
| Material | Graphene Sandwich |
| Areal density [kg/m²] | $7.4\times10^{-7}$ |
| Radius [m] | 183 |
| Mass [g] | 81.6 |

## 2.13 Assembly in Space 100 m KG/ 10× Space Station

The assembly of large structures in space is a challenging task. The largest structure assembled in space to date is the international space station ISS. However, advances in on-orbit manufacturing might soon change the state of the art. The US Company "Made in Space" has announced end of 2015 that it will demonstrate the in-space manufacturing of large booms [96]. The project has been selected by NASA in the context of the Tipping Point Technologies Program. Expected truss lengths are up to 7 km and ring structures with a diameter of 555 m.

Furthermore, Tethers Unlimited is currently developing a system for manufacturing large truss structures in space [97]–[100]. A first demonstrator on a 6 unit CubeSat platform is planned to be launched until 2017 [97]. Its objective is to mature the technology to TRL 4-5.

To summarize, there are serious attempts underway to construct large truss and boom structures in space that have the potential for constructing kilometre-size structures in-situ. The technology has reached a prototypical level and is soon going to be tested in space.

## 2.14 Mass Reduction of Laser Beam Assembly

One of the elements of the laser beam assembly that heavily contributes to the mass of the beam system is primary power generation. The most obvious technology is photovoltaic cells. We leverage on current advances in photovoltaic cells such as organic cells that have a rather low efficiency of <10% but an extremely low areal mass compared to high-efficiency solar cells. Latest developments resulted in a specific-mass 0.1 kg/kW for a terrestrial application [101]. We expect that further advances in photovoltaic cell technology will lead to further increase in mass-specific power output. One disadvantage of organic solar cells is their vulnerability to space radiation. However, degradation issues have been addressed in order to adapt organic cells to space environment [102]–[104].

## 2.15 Beam Assembly Cooling

Heat on the order of several dozens of MW has to be rejected via radiators. High performance radiators are crucial for space exploration, in particular for nuclear-electric propulsion systems. Hence, increasing the mass-specific heat rejection rate has been one of the main objectives of space radiator research. For example, radiators using graphene are expected to reduce the mass by 80-90% with the same heat rejection rate, as it conducts heat five times better than a carbon fibre-reinforced composite sheet [51].

## 2.16 Pointing of Beam Segment at Launch

The pointing requirements for the laser aperture can be calculated by basic trigonometry. There are two effects that affect pointing:

- Laser diffraction, which increases the spot size and decreases the power density of the laser beam on the sail.
- Jitter: Vibrations usually induced by equipment on a spacecraft such as reaction wheels.

Both aspects have been dealt with in the existing laser sail literature such as [2] for diffraction and [105]

## 2.17 Laser Beam Delivery

Laser beam divergence becomes an issue at interstellar distances as too much laser power will eventually spill over the sail and cannot be used for propulsion anymore. While the lens system used to steer and focus the laser beam plays an important role, as discussed in the next subsections, the beam itself can be manipulated to decrease the divergence. Specifically, instead of using a Gaussian beam, which is produced by most laser systems, the beam can be converted to an almost diffraction-less Bessel beam, at least over a limited distance [106][107]. Ideally, one would create a zeroth order Bessel beam using, for example, an Axicon. This beam would propagate with diffraction over a certain distance before returning back to a Gaussian profile. The maximum Bessel-like propagation distance is given by the equation $z_{max} = R/tan\theta = R^2 2\pi/\lambda 2.405$, where $R$ is the aperture radius of the laser emitting system, $\lambda$ the laser wavelength and the factor 2.405 is given by the location of the first root of the Bessel function. The advantage gained by such a system is illustrated in Figure 7. Since an Axicon is a conical prism optical element, however, there are some issues and losses associated with it. For the application in an interstellar lasersail system, the Axicon would have to be big, which in turn makes it heavy using standard manufacturing techniques. Moreover, exact pointing is required. Finally, the optical losses through the Axicon have to be traded off against losses caused by light spilling over the lasersail to choose the optimal solution.

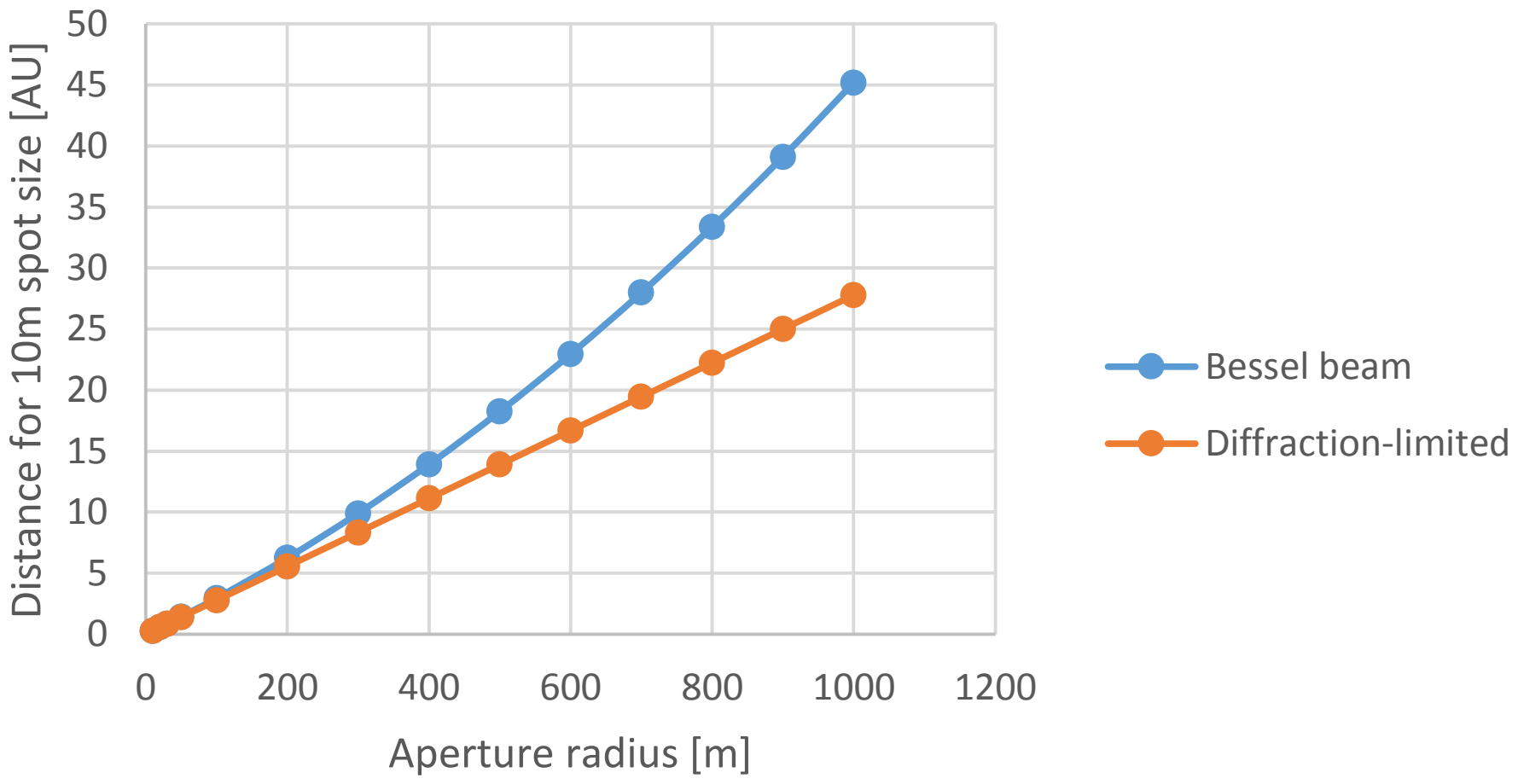

*Figure 7: Focusing capabilities of a purely Gaussian vs. a Bessel-Gaussian beam.*

## 2.18 Lens Configuration

In his 1989 paper [52] Geoffrey Landis looked at the idea of a segmented lens configuration as illustrated in the Figure 8. This means that there would be multiple lenses spread out between the laser and the spacecraft that will facilitate enhanced collimation and pointing accuracy. The table below shows some examples of the beam segmentation architecture and the various options for how it could be staggered.

For the conventional Fresnel lens-based architectures that were originally studied by Robert Forward, high mass ratios are avoided by using a stationary energy source. But this requires a large lens and the high lens/target distance results in an extremely high optical magnification, which presents alignment and positioning difficulties. Some advantage can be gained by using a lightsail reflective in a shorter wavelength regime. The method of lens segmentation however is to use many intermediate lenses spaced between the probe and the source. It is acknowledged however that these will still present pointing, positioning, and deployment challenges.

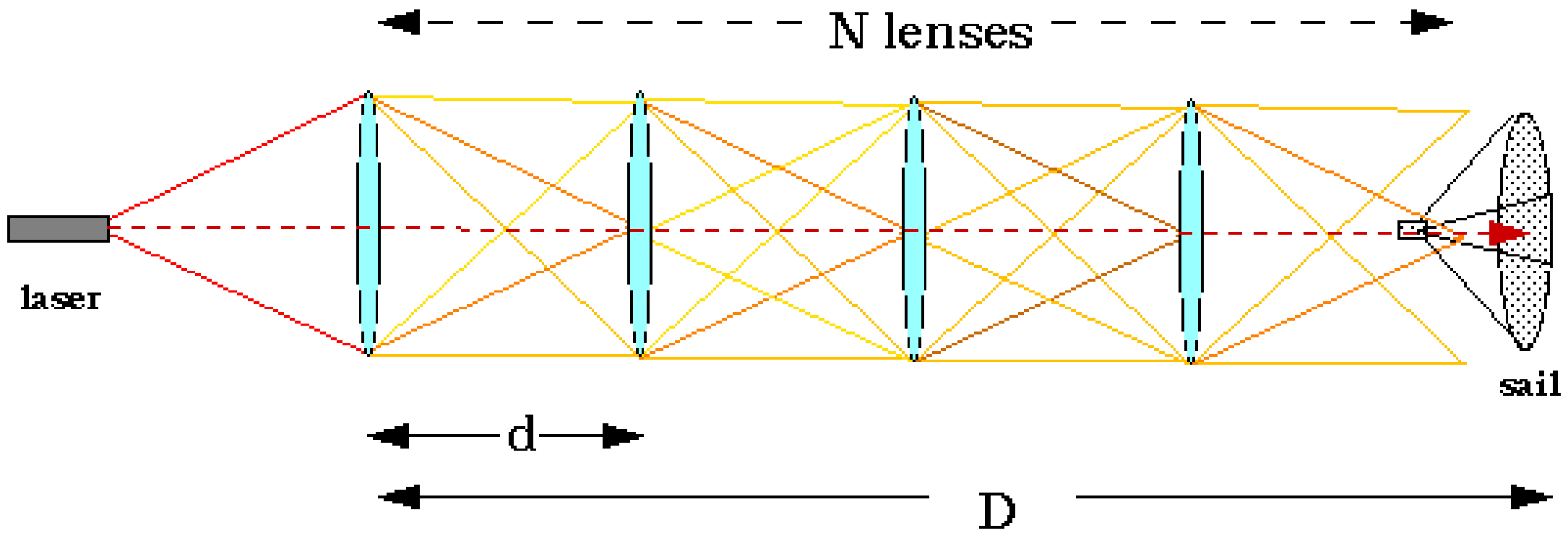

*Figure 8: Use of multiple intermediate lenses to propel a light sail.*

*Table 0-10: Example scaling of Robert Forward Fresnel Lens via Landis Segmentation Scheme*

| Total System Mass (tons) | Number of Lenses | Individual Lens Mass (tons) |
|---|---|---|
| 560,000 | 1 | 560,000 |
| 112,000 | 5 | 22,400 |
| 56,000 | 10 | 5,600 |
| 22,400 | 25 | 896 |
| 11,200 | 50 | 224 |
| 5,600 | 100 | 56 |
| 1,120 | 500 | 2.24 |
| 560 | 1,000 | 0.56 |

Examining this type of segmented architecture, we estimate that the interstellar mission discussed in this report can be achieved with multiple $N$ lenses instead of a single giant lens, and that the mission could be completed with 6 lenses each with a 5 km diameter spread out to around 70 AU.

## 2.18.1 Distributed mirror/lens system

To lower the overall mass and simplify the deployment, instead of using solid lenses, mirrors and other optical elements (e.g. an Axicon) a distributed or inflatable system can be imagined. The orbital rainbow study [108] and DARPA's Membrane Optic Imager Real-Time Exploration (MOIRE) are examples of such concepts. One class of such systems for refractive optics can be described by the "Bruggeman effective medium" approximation. Instead of a space filling medium, particles in a certain size range, with certain shapes and with a certain refractive index are used (e.g. ice crystals) to fill the volume of a prescribed shape by some amount (fill fraction). The changes in refractive index between the background medium (in our case the vacuum of space) and the particles causes refraction to occur with light beams that pass through it. The collective shape can be enforced using a laser trapping system or an inflatable structure. The focal length of such an effective medium can be approximated as $f = 2R/(n_e - 1)$, where $R$ is the aperture radius and $(n_e - 1) = 3/2F(n^2 - 1)/(n^2 + 2)$ is the effective index of refraction, with $F$ being the fill fraction and $n$ being the refractive index of the particles. It should be noted that these equations are derived from more general theories describing the propagation of electromagnetic waves through an effective medium and that an inherent assumption is that the refractive index of space is equal to one. A conceptual sketch showing how such a refractive optic would work is shown in Figure 9.

Some advantages and disadvantages of such a system are listed in Table 11 [108]. The precise light propagation properties and efficiencies of such an effective medium need to be studied further to assess their feasibility.

*Table 11: Advantages and disadvantages of a large distributed optical systems*

| Advantages | Disadvantages |
|---|---|
| 1. Less sensitive to degradation<br>2. Easier to maintain (repair of hull, readjustment of laser trap and injection of new particles)<br>3. Possible use of in-situ resources (e.g. water)<br>4. Significant mass and cost savings | 1. Danger of orbital debris creation by lost particles<br>2. Coherence of very large distributed structures is challenging<br>3. Precise positioning is challenging<br>4. Light loss due to particles not being completely space-filling |

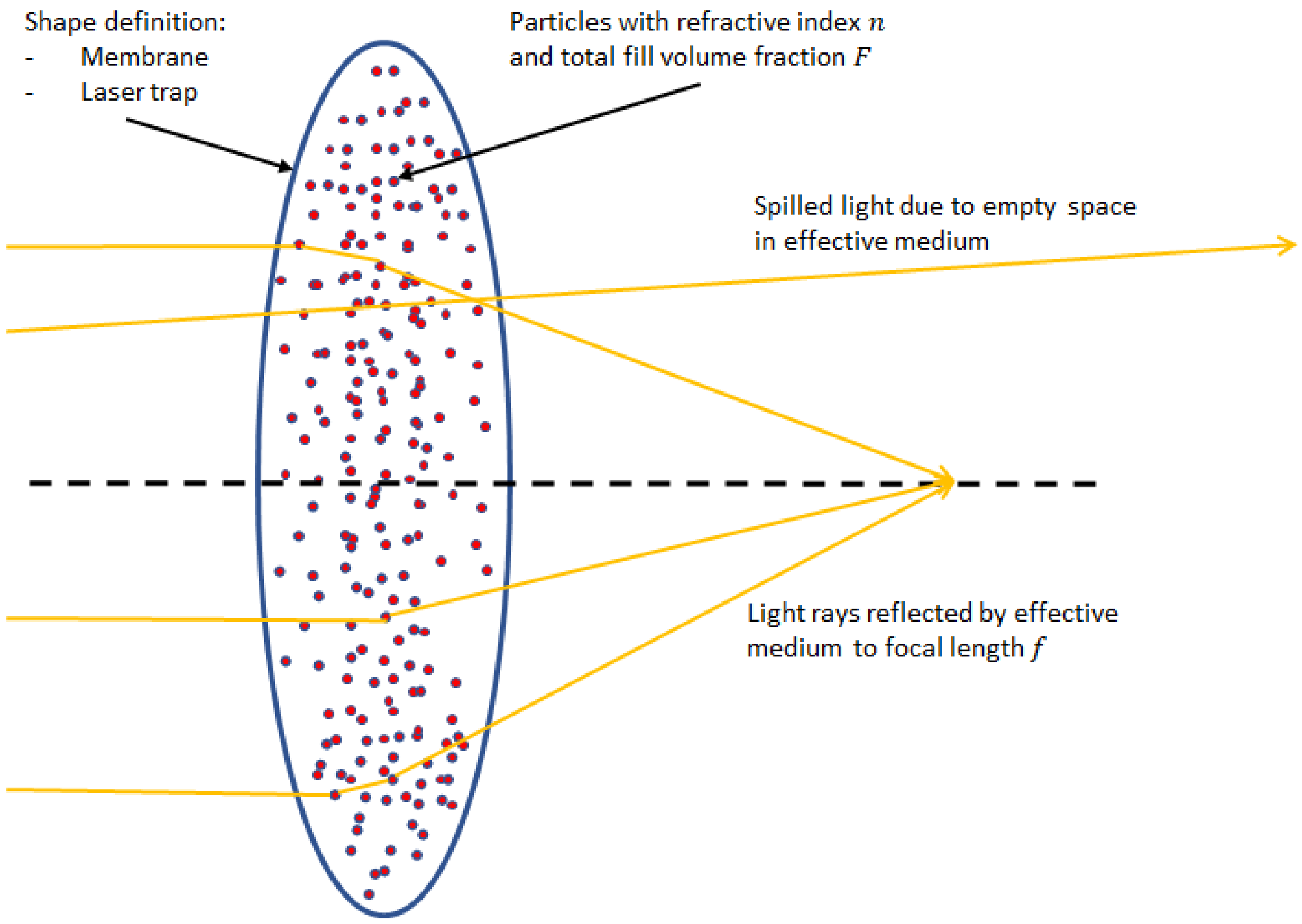

*Figure 9: Conceptual design of a refractive optic using the principle of an effective medium*

## 2.19 Conversion Efficiencies of Beamer

According to Lubin [5] current fibre-fed lasers can have efficiencies near 40%. For example, the DARPA Excalibur program's laser has a specific power of 5kg/kW with prospects of reaching 1 kg/kW near term.

Brashears et al. further assume that laser efficiency will increase to 70% with the specific mass decreasing to 0.1 kg/kW in 10-20 years.

## 2.20 Cost of Mission

***Fixed cost:***

- Development & test of components (sail, communication, energy generation & storage, navigation)
- Developing of specific manufacturing technologies (e.g. in-space manufacturing of large tethers or bubble-blow-up-technologies or mylar-unroll-technologies for the sails) (more needed if many probes should deployed)

***Variable cost:***

- Transport cost of sail & components to space
- Possible in space final assembly of the sail

Variable cost depending on the number of sails, lenses used and the power requirements of the lasers (relating to the space solar power voltaic system size).

**Used NASA Advanced Mission Cost Model and known values to estimate cost of spacecraft development** [109].

*Assumptions:*

- Estimated payload, bus and sail separately
- Wrap factor in spacecraft cost model includes: Adv. Development, Phase A conceptual studies, Phase B definition studies, based on space shuttle program (worst case?)
- Assumed unmanned, exploratory science mission
- Final margin includes: operations capability development, program management and integration, program support wraps, based on space shuttle program (worst case?)
- Assume at least 3 spacecraft and 2 sails are required for development, test and launch
- Technology readiness in 2030
- AMCM includes material, personal, risk, program, development cost
- Does NOT include launch cost,

The summary of the development cost is shown in Table 0-12. It results in a total development cost of about 11M$, including production cost of the first spacecraft units.

***Table 0-12: Development cost for interstellar probe***

| | **Mio US-$** |
|---|---|
| **Total cost** | 11.28 |
| **Wrap Margin** | 1.38 |
| **Total cost + Margin** | 15.54 |
| **Inflation correction from 1990 to 2015** | **28.82** |
| **Phase A and B** | 0.72 |

*Table 0-13: Detailed cost breakdown*

| | Mass | Quantity | Specification | Initial operational Capability | Block | Difficulty | Complexity | Wrap | Risk score | Reserve | Total Mio US-$ |
|---|---|---|---|---|---|---|---|---|---|---|---|
| ***Concept 2: Alternative power source*** | | | | | | | | | | | |
| **Payload [g]** | | | | | | | | | | | |
| Sensor package | 4 | | | | | | | | | | |
| nano FPGA-based OBDH | 30 | | | | | | | | | | |
| Payload mass | 0.034 | 3 | 2.2 | 2030 | 2 | -1 | 1 | 1.122 | 5 | 1.25 | 0.484514529 |
| **Power** | | | | | | | | | | | |
| Graphene supercapacitors | 25 | | | | | | | | | | |
| Pico RTG | 40 | | | | | | | | | | |
| **Structure** | 5 | | | | | | | | | | |
| **Communications (laser)** | 26 | | | | | | | | | | |
| **Startracker / camera + telescope** | 50 | | | | | | | | | | |
| Radiation protection (Polyethylene) | 20 | | | | | | | | | | |
| ***Bus mass*** | 0.166 | 3 | 2.2 | 2030 | 2 | -1 | 1.5 | 1.122 | 8 | 1.4 | 2.319459486 |
| **Sail** | 80 | | | | | | | | | | |
| | 0.08 | 2 | 2.2 | 2030 | 1 | 2 | 2 | 1.122 | 13 | 1.6 | 8.477166798 |
| ***Total mass [g]*** | 280 | | | | | | | | | | |
| | | | | | | | | | | | **Mio US-$** |
| | | | | | | | | | | **Total cost** | 11.28114081 |
| | | | | | | | | | | **Wrap Margin** | 1.378 |
| | | | | | | | | | | **Total cost + Margin** | 15.54541204 |
| | | | | | | | | **Inflation correctionfrom 1990 to 2015** | | | 28.82032491 |
| | | | | | | | | | | **Phase A and B** | 0.720508123 |

For any future analysis, the overall mission cost using projected transportation should take account of cost decreases (as projected by SpaceX rocket reuse scenario) and this will need to be considered.

## 2.21 National and International Policies/Treaties on use of Directed Energy Devices in Space

It is necessary to conduct a brief survey of the key laws, conventions and treaties which will be relevant to such an innovative laser sail mission given the use of directed energy sources. We do flag up for the attention of the *Breakthrough Initiative* that is not clear to this team whether lasers are considered to be conventional weapons or not. The Geneva Convention bans the building of lasers under a special conventional weapons treaty but we are unsure of the clear definitions which are covered. Another point to make is that some of the recent activities on treaties could potentially be in favour of such a mission, such as the recent addition of rights to utilize space resources and the right to keep control of launched objects. The following are some of the relevant articles which this team feels are relevant to such a mission as proposed for Andromeda.

**Outer Space Treaty**

- Article 8: The State that launches a space object retains jurisdiction and control over that object
- Article 7: The launching state is absolutely liable for damage caused on Earth's surface or to aircraft in flight; if the damage is caused elsewhere (e.g., in space), the launching state is liable only if the damage is due to its fault or the fault of persons for whom it is responsible
- Article 4: Limits the use of the Moon and other celestial bodies to peaceful purposes and expressly prohibits their use for testing weapons of any kind, conducting military manoeuvres, or establishing military bases, installations, and fortifications. However, the Treaty does not prohibit the placement of conventional weapons in orbit. → *Lasers seem to be conventional weapons*
- Agreement Governing the Activities of States on the Moon and Other Celestial Bodies:
    - Exploration and use of the Moon shall be carried out for the benefit and in the interest of all countries, and due regard shall be paid to the interests of present and future generations and to the need to promote higher standards of living and conditions of economic and social progress and development in accordance with the U.N. charter.
    - States Parties bear international responsibility for national activities on the Moon, whether by governmental or non-governmental entities. Activities of non-governmental entities must take place only under the authority and continuing supervision of the appropriate State Party.

- All space vehicles, equipment, facilities, etc. shall be open to other States Parties so all States Parties may assure themselves that activities of others are in conformance with this agreement. Procedures are established for resolving differences.

**Space Resource Exploration and Utilization Act of 2015**

- Sec 2:
  - Discourage government barriers to the development of economically viable, safe, and stable industries for the exploration and utilization of space resources in manners consistent with the existing international obligations of the United States
  - Promote the right of U.S. commercial entities to explore outer space and utilize space resources, in accordance with such obligations, free from harmful interference, and to transfer or sell such resources.

Defines "space resource" as a natural resource of any kind found in place in outer space. → *Solar power a space resource?*

**Outer Space Treaty and Weapons in Space**

We argue that a laser is not a non-discriminatory weapon, as it is targeted, therefore no weapon of mass destruction (WMD). Deployment in Earth orbit or the "outer void space" of non-prohibited weapon types is allowed under the terms of the Outer Space treaty. Deployment on the moon and other celestial bodies is prohibited.

*Interpretation:* Lasers are no WMD (even so they can be deemed as weapons) and we do not plan to deploy on them on the Moon or any other celestial body but in Earth orbit; so according to the paper by Minero (2008) it would be legal to install Lasers in orbit under the Space Treaty.

See also information on the definition and discussion of conventional/discriminatory weapons [55].

# 3. CONCEPT DESIGN FOR ANDROMEDA PROBE

Here we briefly lay out a concept for a probe design based on all the considerations discussed above. This concept is not the final design, and a greater level of sub-system fidelity is required to fully sketch out the performance. A graphic of the possible probe design is shown below as a configuration layout.

Figure 11 shows an overview of the components of the spacecraft. The probe is fitted with a camera (star tracker) and has a foldable aperture. It can be folded onto the main plate structure of the spacecraft. A CMOS camera is integrated into the spacecraft structure. A nuclear battery is positioned as a circular object on the back. The excess heat is used for heating the spacecraft components. Hence there are very thin copper circuits running from the battery to the critical components.

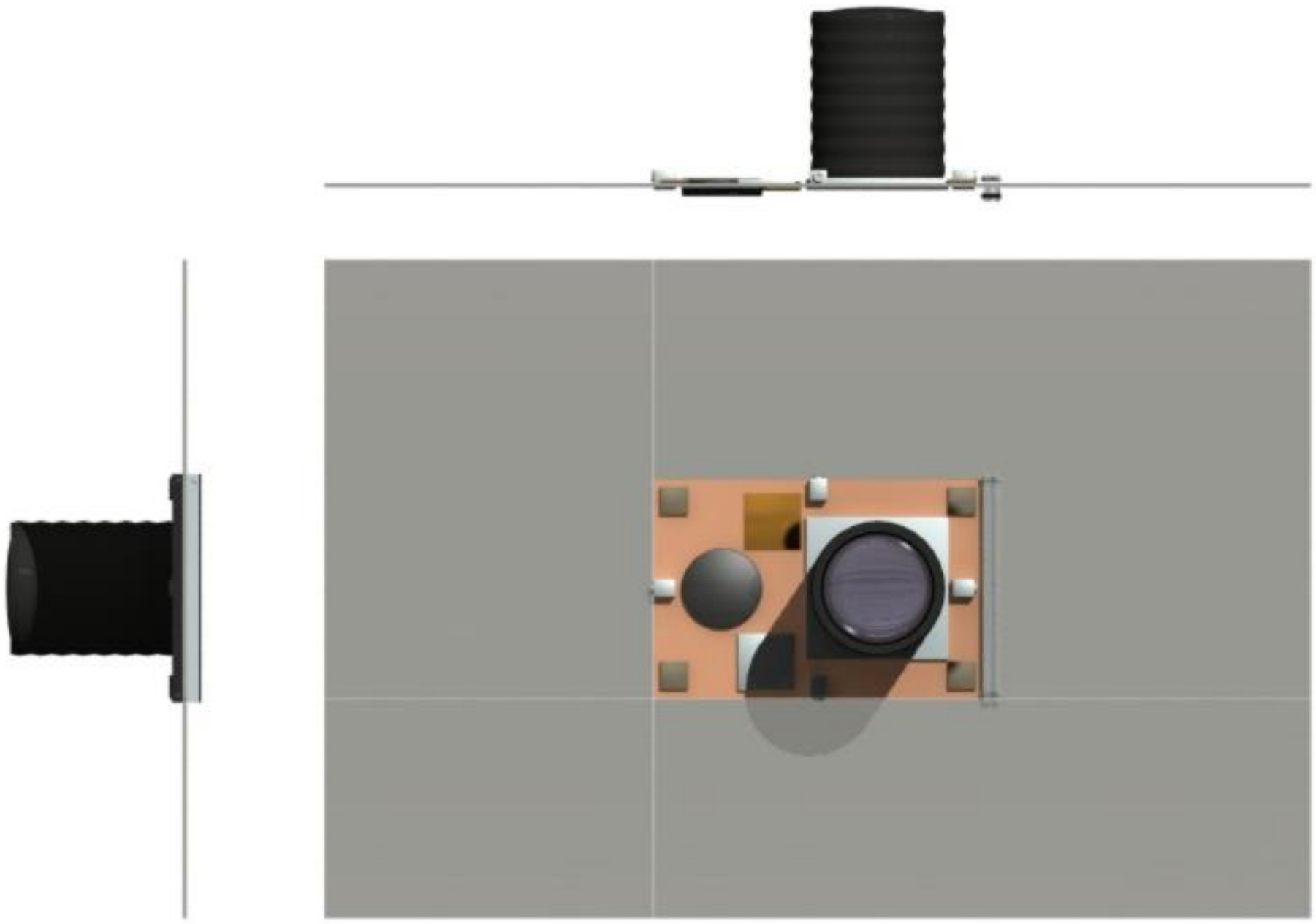

*Figure 10: Orthographic Layout Drawing for Interstellar Probe (Credit: Adrian Mann)*

Graphene based supercapacitors are positioned at the side of the nuclear battery. The battery then loads the supercapacitors, and then they supply power to selective components such as the antenna, OBDH etc. The spacecraft structure has distributed MEMS sensors such as magnetometers etc. The antenna is a phased array, which is folded at the bottom of the spacecraft, approximately 1 m on each edge when unfolded.

A graphene whipple shield is used at the front of the spacecraft for protecting the spacecraft from the interstellar matter that is incoming. The whipple shield consists of multiple layers of shield material. Each laser is intended to extract energy from the incoming particle. Part of the energy is released via radiation, when a plasma is generated upon impact. The plasma cools down immediately, releasing its energy via radiation during its transit to the next layer in the shield. Hence, the multiple layers of the shield are more effective than a bulk shield. Alternative materials such as Beryllium are also possible. The shield is approximately 1 mm thick and 5 cm to 10 cm wide, depending on the width of the spacecraft. The spacecraft flies in the direction of the graphene shield, once it has completed its acceleration phase. Note that the graphene sail provides additional shielding before it is eroded away.

The spacecraft has actuators. These are MEMS actuators (gyros) that are distributed on the spacecraft for providing attitude control. There are also MEMS propulsion units for desaturating gyro's occasionally.
During the acceleration phase, the spacecraft is attached to the sail via graphene wires. The sail has optomechanical elements for steering. The sail has furthermore a slightly conic shape in order to self-adjust to the laser beam maximum

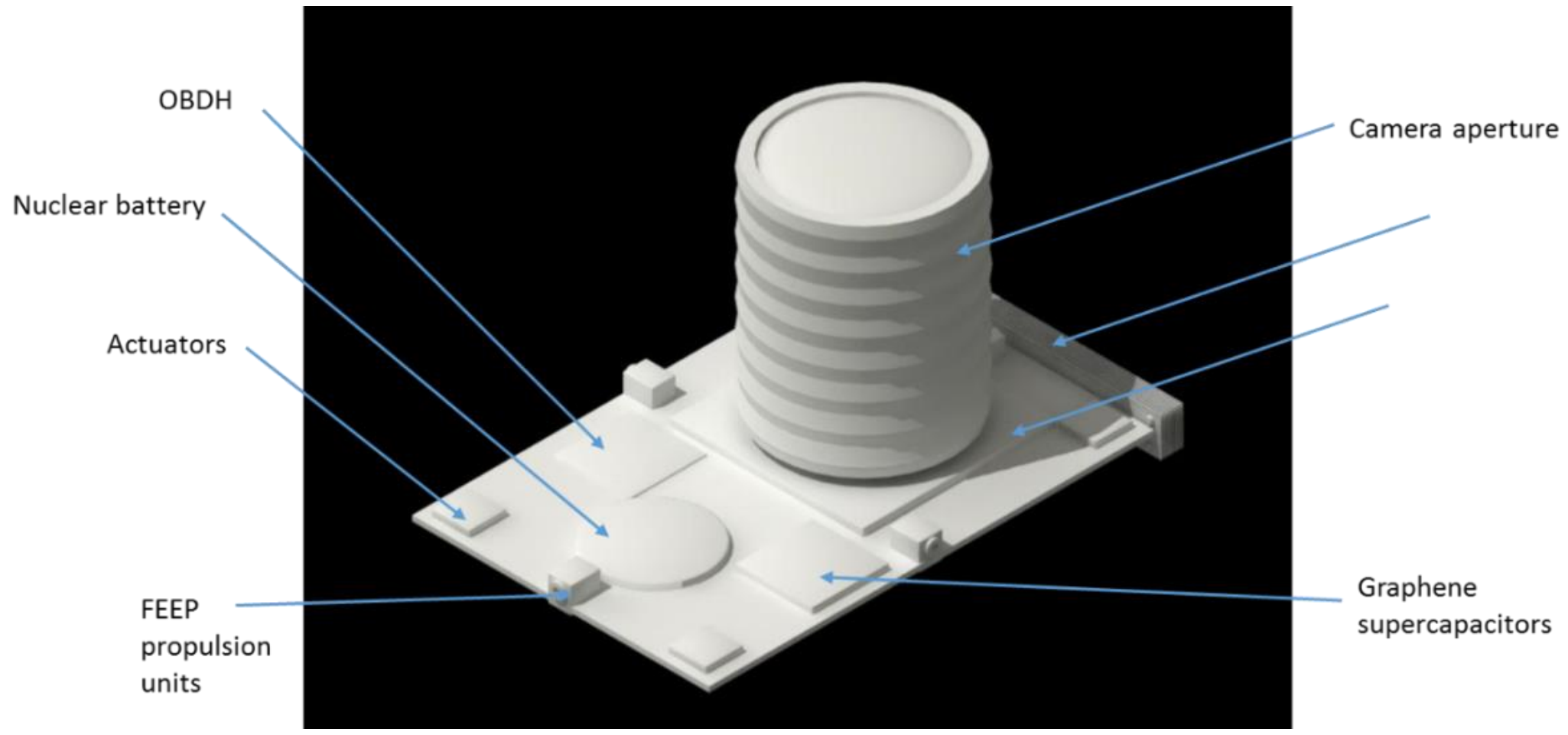

*Figure 11: Overview of spacecraft components (Credit: Adrian Mann)*

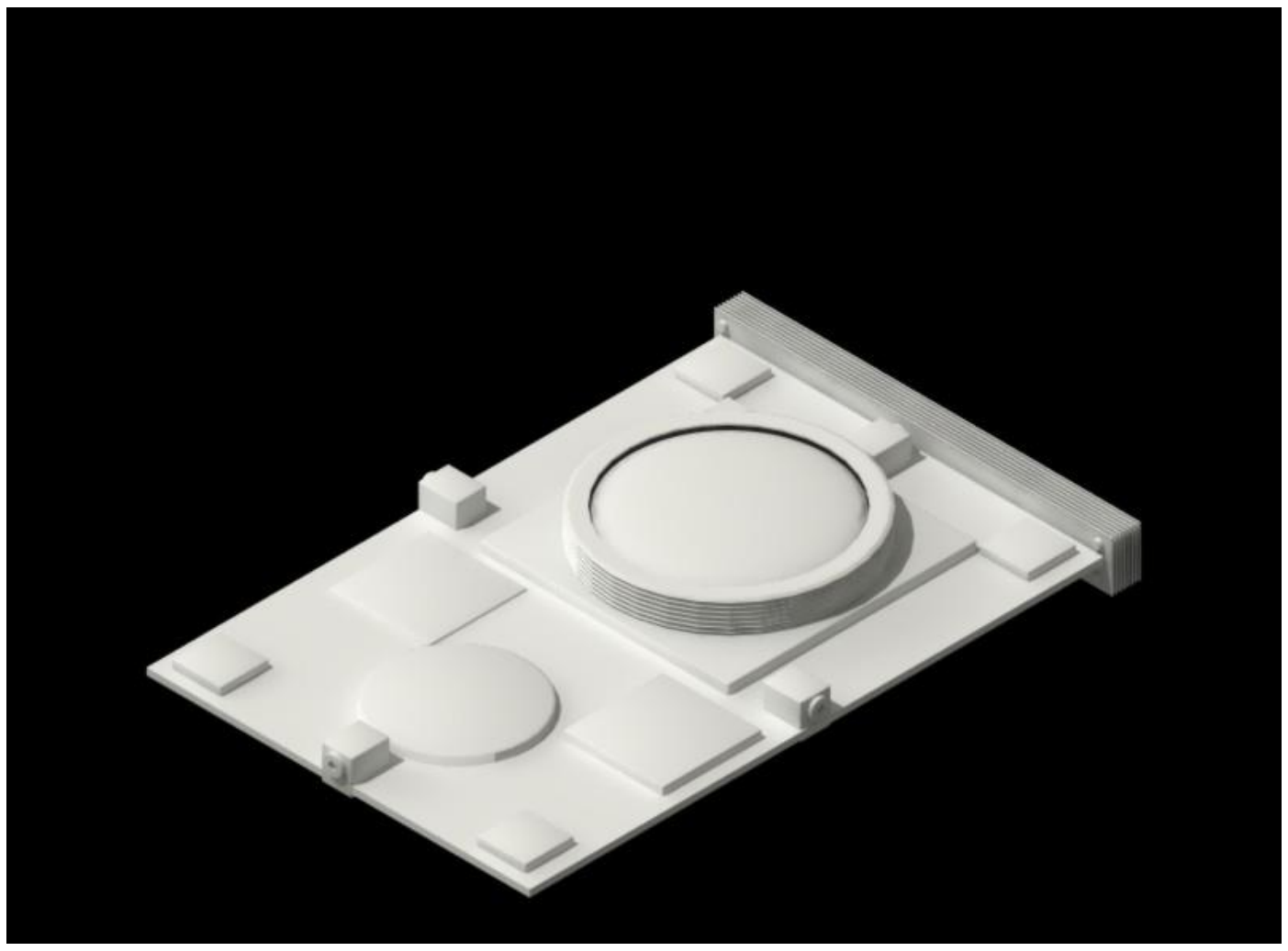

*Figure 12: Spacecraft with folded camera aperture for protecting against interstellar matter (Credit: Adrian Mann)*

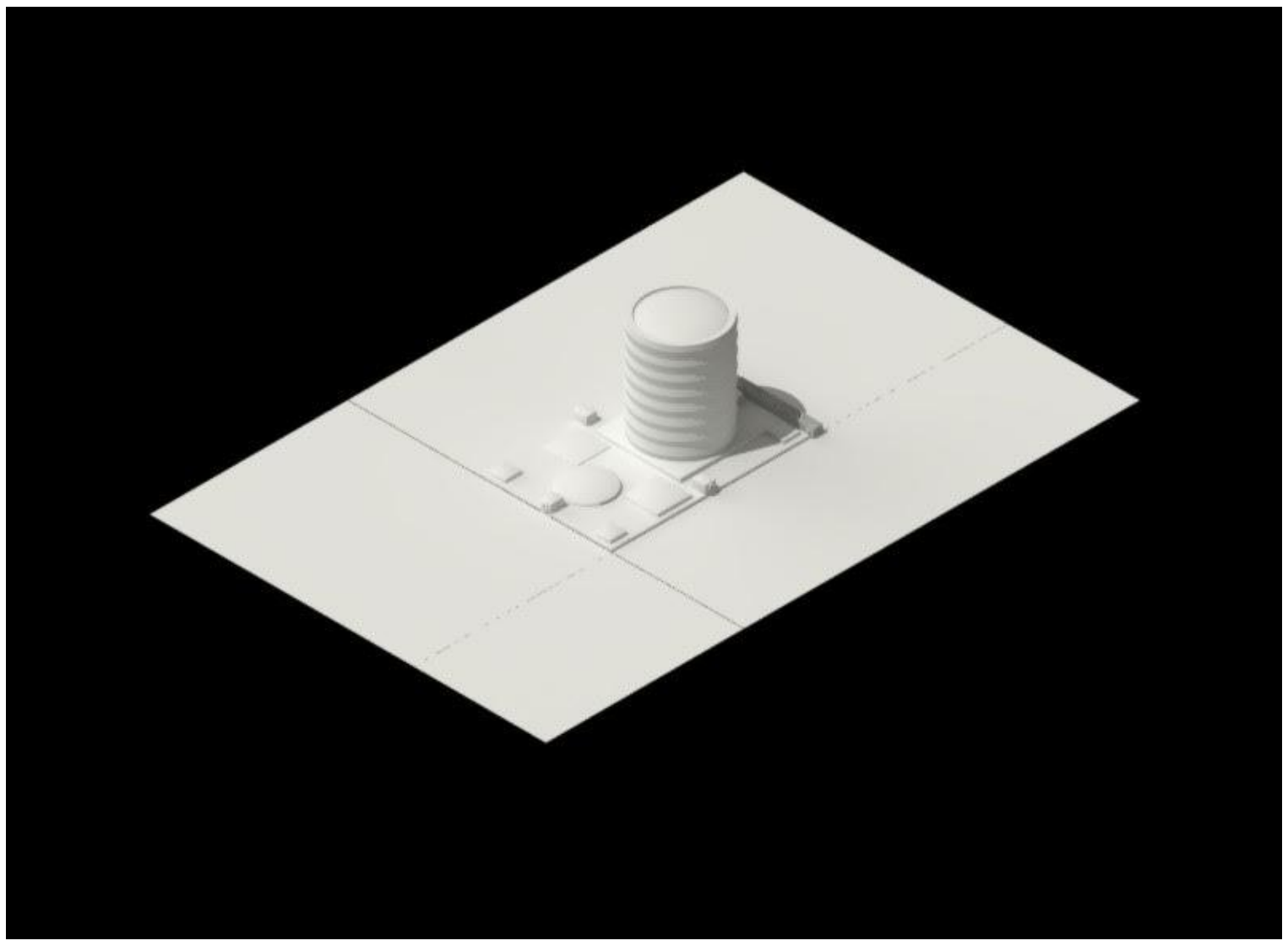

*Figure 13: Spacecraft with deployed antenna panels (Credit: Adrian Mann)*

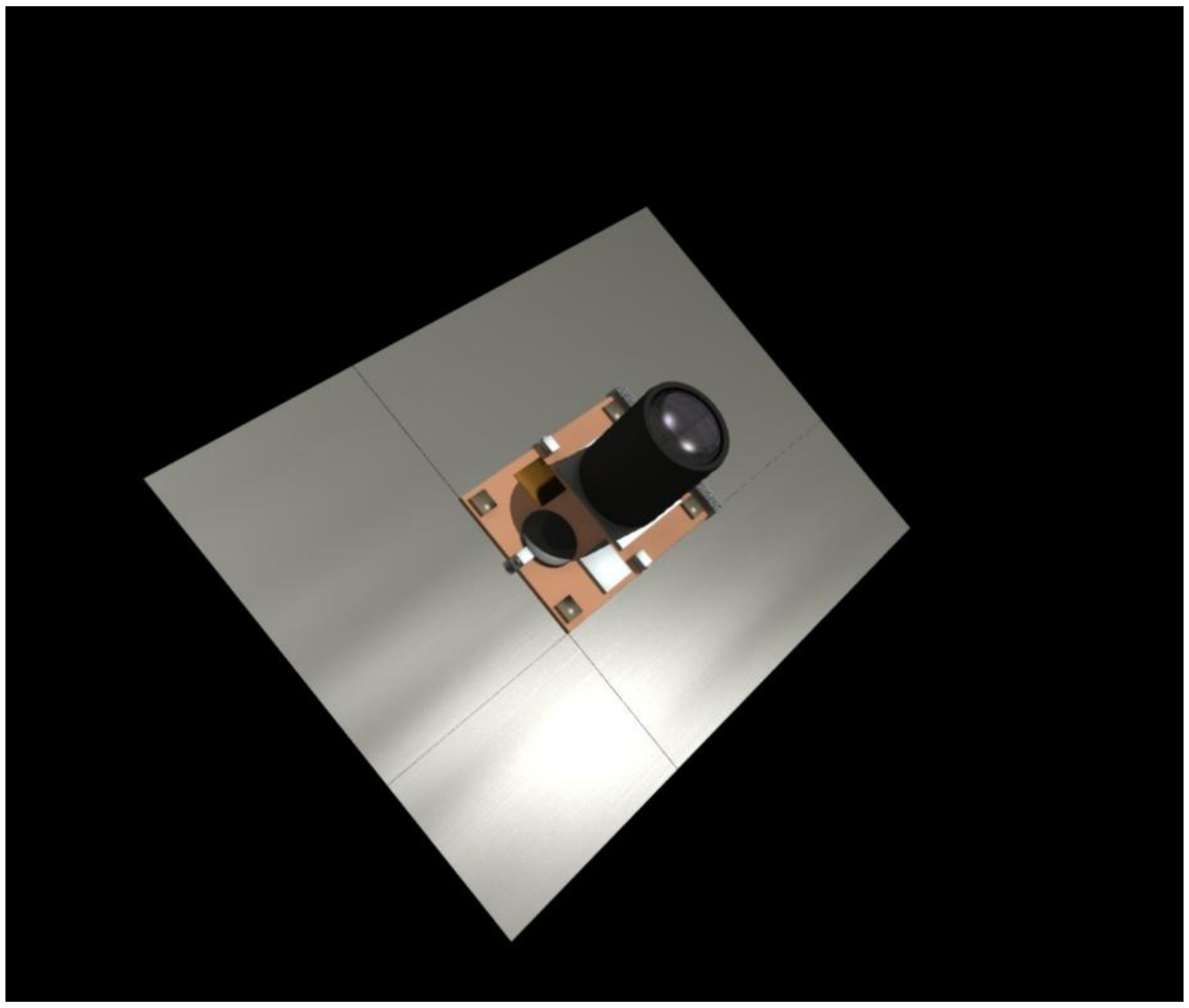

*Figure 14: Spacecraft with deployed antenna panels (Credit: Adrian Mann)*

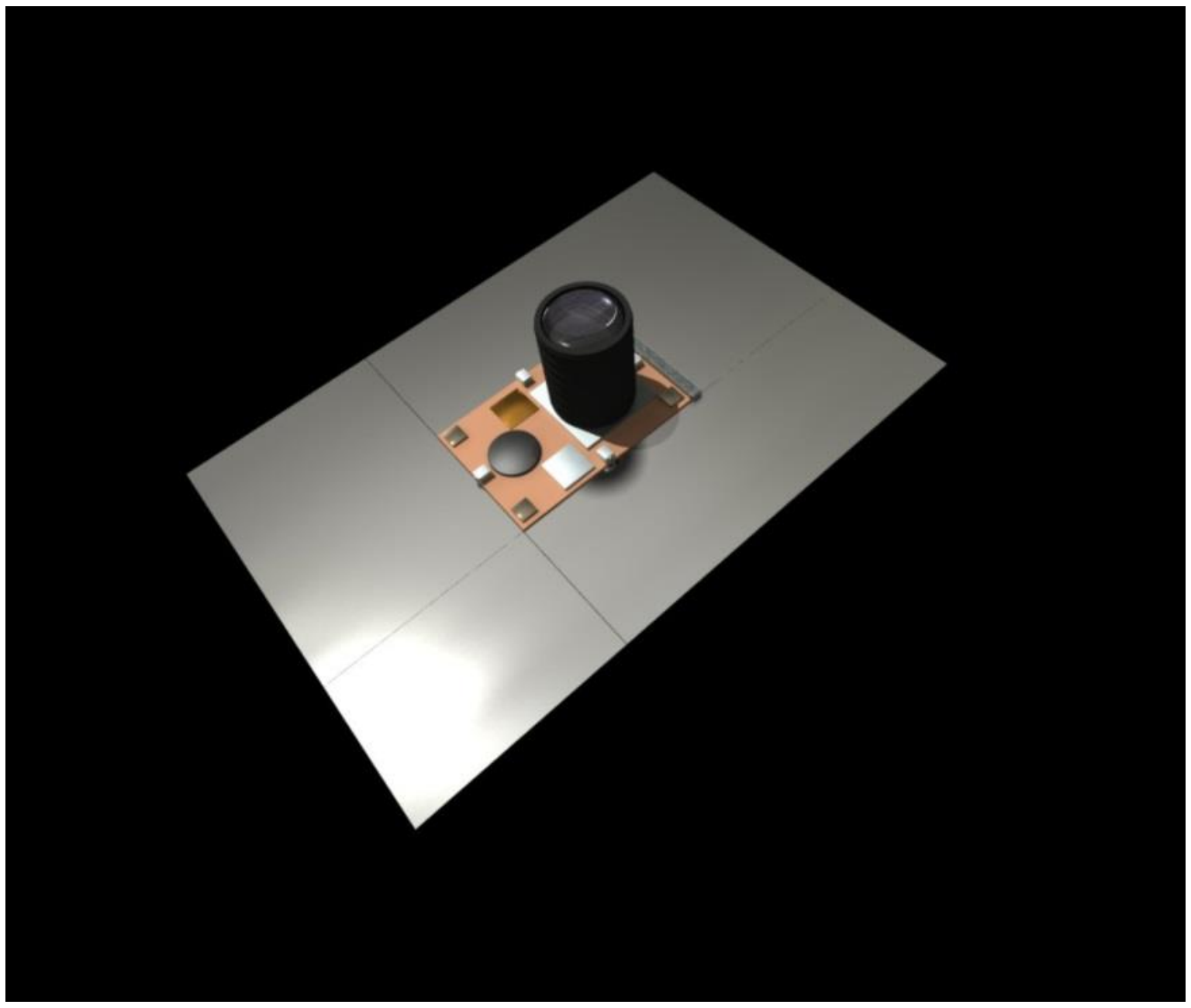

*Figure 15: Illustration of Interstellar Probe (Credit: Adrian Mann)*

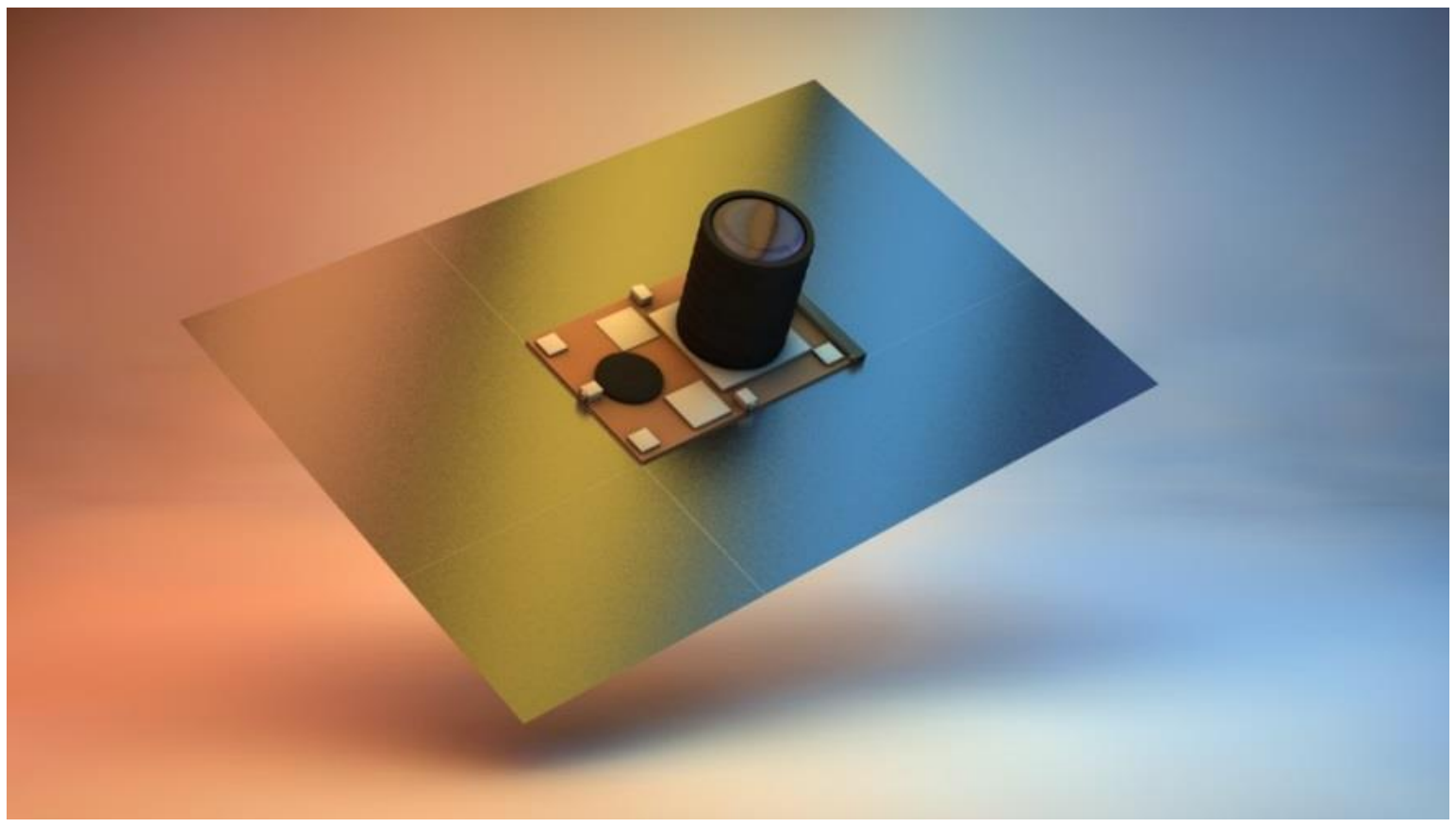

*Figure 16: Spacecraft with deployed antenna panels (Credit: Adrian Mann)*

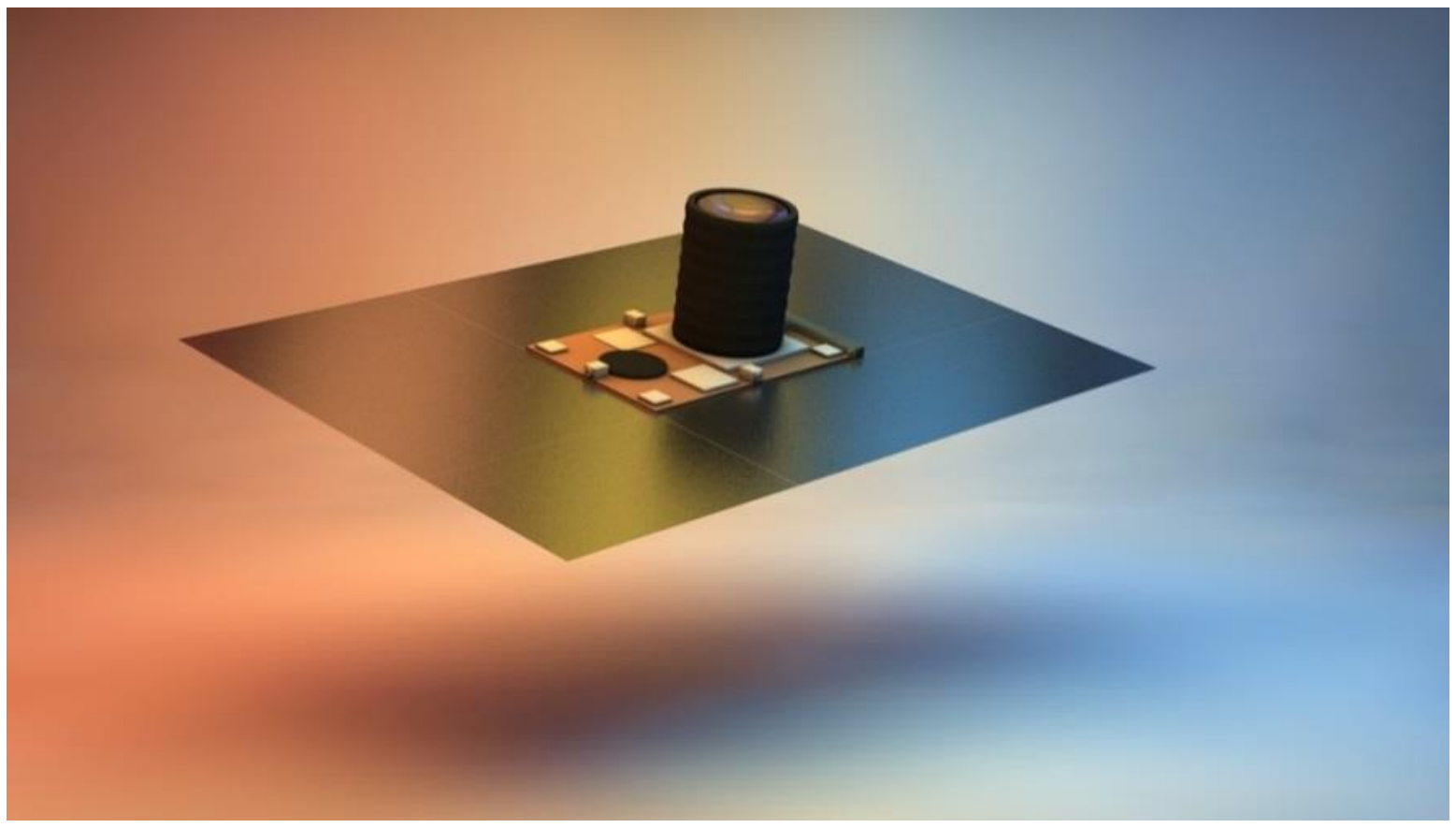

*Figure 17: Spacecraft with deployed antenna panels (Credit: Adrian Mann)*

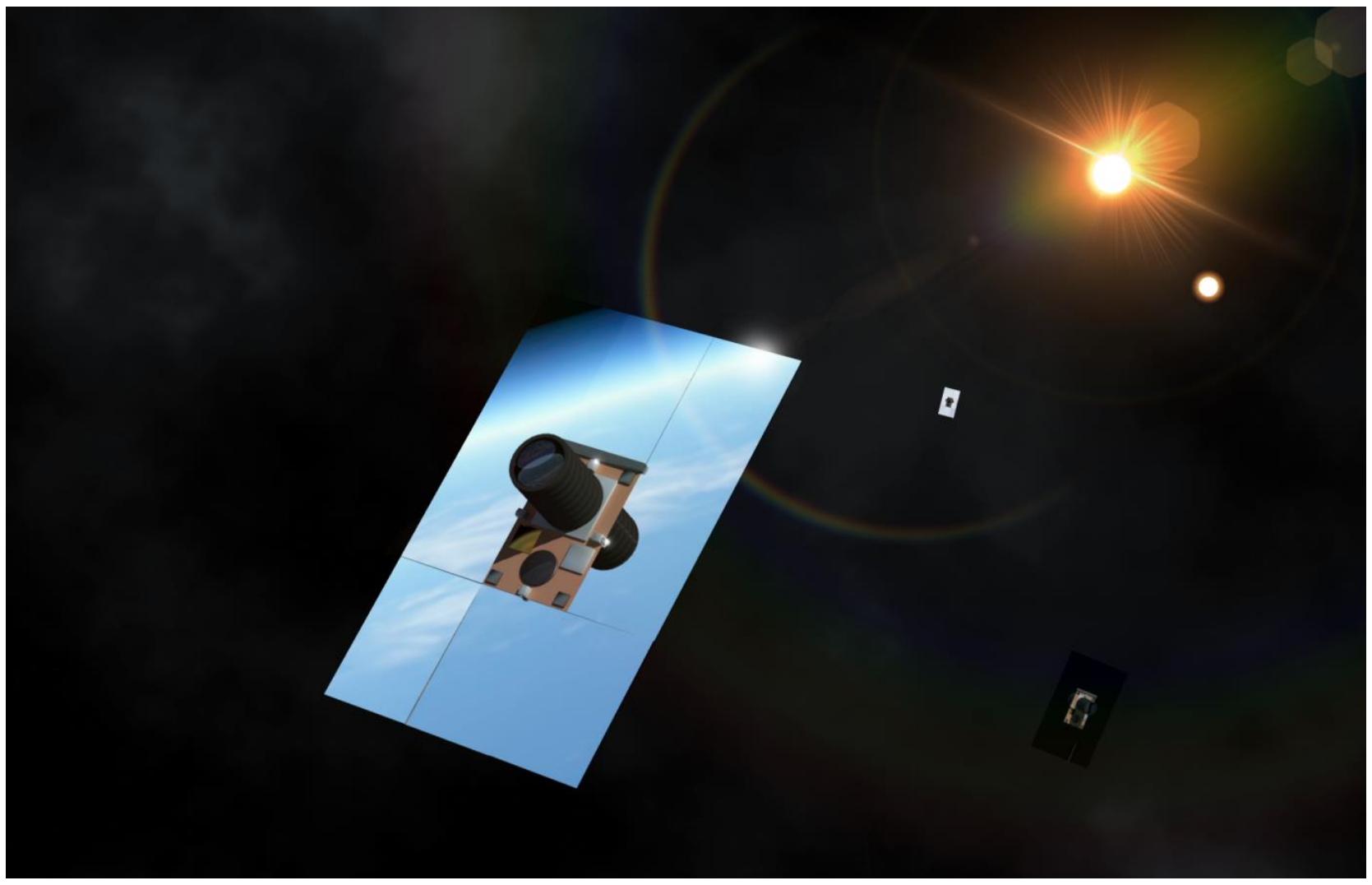

*Figure 18: Spacecraft in flight (Credit: Adrian Mann)*

**Possible architectures with mass budget**

The mass budget for the spacecraft is shown in Table 3-1. Most of the component masses seem to be exceptionally low. However, we believe that advances in miniaturization will soon attain a level where these low masses are actually feasible. For several components, such as the OBDH, we have taken values for existing components. The total mass is 280g without the mass margin. If a 50% margin is added, this mass increases to 420g. However, in the rest of the report, we use the lower 280g value, as we think that the component masses we assumed are rather conservative with respect to future advances in electronics.

*Table 3-1: Mass budget for spacecraft*

| | **Mass [g]** |
|---|---|
| **Payload** | |
| Sensor package | 4 |
| Nano FPGA-based OBDH | 30 |
| **Power** | |
| Graphene supercapacitors | 25 |
| Nuclear battery | 40 |
| **Structure** | 5 |
| **Communications RF** | 26 |
| **ADCS** | 10 |
| **Startracker / camera + telescope** | 20 |
| **Radiation protection (Polyethylene)** | 20 |
| **Interstellar dust protection** | 20 |
| **Bus mass** | 200 |
| **Sail** | 80 |
| **Total mass** | 280 |
| **50% margin** | 140 |
| **Mass with margin** | 420 |

The data below shows the output from an in-house laser-sail physics code written in Fortran 95 which calculates the mission profile for a desired scenario. The calculations show that the probe will accelerate at 22 $m/s^2$ for 16 days until it reaches a distance of 140.5 AU, hitting a cruise velocity of 30,408 km/s or 0.1c getting to its target destination Alpha Centauri at 4.3 light years after 42 years of flight. The total laser power required to propel this vehicle is around 1.12 GW assuming a 500 nm wavelength laser.

```
#############################################################
 PROMETHEUS LASER-SAIL PROPULSION CODE
#############################################################
Centauri A
Destination distance (LY, AU):   4.30000019       271926.156
 One way flyby mission
 ============================
 Data for out route Sail
 ============================
 Sail material is Graphene
 Sail geometry is circular
 Sail diameter (m):   366.000000
 Sail mass (kg):  0.280000001
 Sail loading (g/m2):   2.66137440E-03
 Lens distance from Light Source (m, AU):   2.24399983E+12   15.0000000
 Solar irradiance at launch (W/m^2):   6.03709412
 Sail density (g/cm^3):   2.20000005
 Sail thickness (nm):   16.0000000
 Sail areal density (g/m^2):  0.100000001
 Sail Reflectivity:  0.819999993
 Sail Emissivity:   5.99999987E-02
 Sail Absorption Coefficient:  0.135000005
 Sail Temperature:   600.000000
 Acceleration (m/s2):   22.0000000
 Earth Gravities (1/ge):   2.24260950
 Lightness number (m/s^2):   3709.94946
 Laser Wavelength (m):   4.99999999E-07
 Sail Power requirement (GW):   1.12604964
  ============================
 Mission Performance
 ============================
 Cruise velocity (km/s, %c):   30408.9023       10.1433182
 Boost time (Years):   4.38299999E-02
 Boost distance (m, AU, LY):   2.10159419E+13   140.480896       2.22144090E-03
 Beam spot size at boost cut-off (km):   2.52191313E-02
 Cruise distance (m, AU):   4.06591368E+16   271785.688
 Cruise time (Years):   42.3985291
 Total mission distance (m, AU,LY):   4.06801521E+16   271926.156       4.30000019
 Total mission duration (Years):   42.4423599
  ============================================================
 END OF CALCULATION SUMMARY
```

To conclude, extrapolating from current trends in microelectronics and their use in CubeSats and Femtosats, we can assume that an interstellar probe as presented in this section could become feasible within a timeframe of 10 to 20 years.

# 4. I4IS LASER SAIL DEMONSTRATOR MISSION

*The Initiative for Interstellar Studies* has recently begun work on plans for a CubeSat mission into space that deploys the world's first laser-sail. The philosophy of this mission is "***keep it simple, cheap, get to flight readiness quickly and de-risk the physics and engineering"*** by maximum ground validation and certification pre-flight. The current plan is to aim for a launch in the 2017-2020 timeframe in celebration of the 60$^{th}$ anniversary of the Sputnik 1 mission which kickstarted the space race. Our aspiration is to kickstart a new era of space exploration but with laser-sail technology and by demonstrating it is possible.

Our probe design is small, of order 9 $cm^2$ in area for the sail and 1 $cm^2$ in area for the ChipSat that is deployed with it. We estimate that at LEO such a sail could receive around 1 W of power just from solar energy alone, although the Earth albedo and atmospheric drag may reduce this if the launch altitude is not high enough. This power would be sufficient to push a 10s gram probe to 100s m/s. The calculations below show examples for such probes over variations in sail parameters and probe mass.

*Table 4-1: Calculations for 0.16 W ChipSats*

| Reflectivity μ<br>Pressure $P_r$ | 20 g | 30 g | 40 g |
|---|---|---|---|
| $\mu = 1.0$<br>$P_r = 9.053\times10^{-6}$ N/m$^2$ | $\sigma = 200$ kg/m$^2$<br>$a_c = 4.526\times10^{-8}$ m/s$^2$<br>$\lambda = 7.633\times10^{-6}$<br>$v_{esc} = 116.3$ m/s<br>$S_{1min} = 6.9$ km | $\sigma = 300$ kg/m$^2$<br>$a_c = 3.017\times10^{-8}$ m/s$^2$<br>$\lambda = 5.088\times10^{-6}$<br>$v_{esc} = 94.9$ m/s<br>$S_{1min} = 5.7$ km | $\sigma = 400$ kg/m$^2$<br>$a_c = 2.263\times10^{-8}$ m/s$^2$<br>$\lambda = 3.816\times10^{-6}$<br>$v_{esc} = 82.2$ m/s<br>$S_{1min} = 4.9$ km |
| $\mu = 0.8$<br>$P_r = 8.148\times10^{-6}$ N/m$^2$ | $\sigma = 200$ kg/m$^2$<br>$a_c = 4.074\times10^{-8}$ m/s$^2$<br>$\lambda = 6.870\times10^{-6}$<br>$v_{esc} = 110.3$ m/s<br>$S_{1min} = 6.6$ km | $\sigma = 300$ kg/m$^2$<br>$a_c = 2.716\times10^{-8}$ m/s$^2$<br>$\lambda = 4.580\times10^{-6}$<br>$v_{esc} = 90.1$ m/s<br>$S_{1min} = 5.4$ km | $\sigma = 400$ kg/m$^2$<br>$a_c = 2.037\times10^{-8}$ m/s$^2$<br>$\lambda = 3.435\times10^{-6}$<br>$v_{esc} = 78.0$ m/s<br>$S_{1min} = 4.7$ km |
| $\mu = 0.6$<br>$P_r = 7.243\times10^{-6}$ N/m$^2$ | $\sigma = 200$ kg/m$^2$<br>$a_c = 3.621\times10^{-8}$ m/s$^2$<br>$\lambda = 6.107\times10^{-6}$<br>$v_{esc} = 104.0$ m/s<br>$S_{1min} = 6.2$ km | $\sigma = 300$ kg/m$^2$<br>$a_c = 2.414\times10^{-8}$ m/s$^2$<br>$\lambda = 4.071\times10^{-6}$<br>$v_{esc} = 84.9$ m/s<br>$S_{1min} = 5.1$ km | $\sigma = 400$ kg/m$^2$<br>$a_c = 1.811\times10^{-8}$ m/s$^2$<br>$\lambda = 3.053\times10^{-6}$<br>$v_{esc} = 73.6$ m/s<br>$S_{1min} = 4.4$ km |

**Trade Space for WISP Sprites 1 cm$^2$ area with 0.16 W Solar Energy Input**

*Table 4-2: Calculations for 0.54 W ChipSats*

| Reflectivity μ<br>Pressure $P_r$ | 20 g | 30 g | 40 g |
|---|---|---|---|
| $\mu = 1.0$<br>$P_r = 9.053\times10^{-6}$ N/m$^2$ | $\sigma = 50$ kg/m$^2$<br>$a_c = 1.811\times10^{-7}$ m/s$^2$<br>$\lambda = 7.633\times10^{-6}$<br>$v_{esc} = 116.3$ m/s<br>$S_{1min} = 6.9$ km | $\sigma = 75$ kg/m$^2$<br>$a_c = 2.422\times10^{-8}$ m/s$^2$<br>$\lambda = 5.088\times10^{-6}$<br>$v_{esc} = 94.9$ m/s<br>$S_{1min} = 5.7$ km | $\sigma = 100$ kg/m$^2$<br>$a_c = 9.053\times10^{-8}$ m/s$^2$<br>$\lambda = 3.816\times10^{-6}$<br>$v_{esc} = 82.2$ m/s<br>$S_{1min} = 4.9$ km |
| $\mu = 0.8$<br>$P_r = 8.148\times10^{-6}$ N/m$^2$ | $\sigma = 50$ kg/m$^2$<br>$a_c = 1.629\times10^{-7}$ m/s$^2$<br>$\lambda = 6.870\times10^{-6}$<br>$v_{esc} = 110.3$ m/s<br>$S_{1min} = 6.6$ km | $\sigma = 75$ kg/m$^2$<br>$a_c = 1.219\times10^{-7}$ m/s$^2$<br>$\lambda = 4.580\times10^{-6}$<br>$v_{esc} = 90.1$ m/s<br>$S_{1min} = 5.4$ km | $\sigma = 100$ kg/m$^2$<br>$a_c = 8.148\times10^{-8}$ m/s$^2$<br>$\lambda = 3.435\times10^{-6}$<br>$v_{esc} = 78.0$ m/s<br>$S_{1min} = 4.7$ km |
| $\mu = 0.6$<br>$P_r = 7.243\times10^{-6}$ N/m$^2$ | $\sigma = 50$ kg/m$^2$<br>$a_c = 1.449\times10^{-7}$ m/s$^2$<br>$\lambda = 1.940\times10^{-5}$<br>$v_{esc} = 185.4$ m/s<br>$S_{1min} = 11.1$ km | $\sigma = 75$ kg/m$^2$<br>$a_c = 9.657\times10^{-8}$ m/s$^2$<br>$\lambda = 1.628\times10^{-5}$<br>$v_{esc} = 169.9$ m/s<br>$S_{1min} = 10.2$ km | $\sigma = 40$ kg/m$^2$<br>$a_c = 1.811\times10^{-7}$ m/s$^2$<br>$\lambda = 3.053\times10^{-5}$<br>$v_{esc} = 232.6$ m/s<br>$S_{1min} = 13.9$ km |

. **Trade Space for WISP Sprites 4 cm$^2$ area with 0.54 W Solar Energy Input**

*Table 4-3: Calculations for 1.22 W ChipSats*

| Reflectivity μ<br>Pressure $P_r$ | 20 g | 30 g | 40 g |
|---|---|---|---|
| $\mu = 1.0$<br>$P_r = 9.053\times10^{-6}$ N/m$^2$ | $\sigma = 22.2$ kg/m$^2$<br>$a_c = 4.078\times10^{-7}$ m/s$^2$<br>$\lambda = 6.877\times10^{-5}$<br>$v_{esc} = 349.1$ m/s<br>$S_{1min} = 20.9$ km | $\sigma = 33.3$ kg/m$^2$<br>$a_c = 2.719\times10^{-7}$ m/s$^2$<br>$\lambda = 4.585\times10^{-5}$<br>$v_{esc} = 285.1$ m/s<br>$S_{1min} = 17.1$ km | $\sigma = 44.4$ kg/m$^2$<br>$a_c = 2.039\times10^{-7}$ m/s$^2$<br>$\lambda = 3.438\times10^{-5}$<br>$v_{esc} = 246.8$ m/s<br>$S_{1min} = 14.8$ km |
| $\mu = 0.8$<br>$P_r = 8.148\times10^{-6}$ N/m$^2$ | $\sigma = 22.2$ kg/m$^2$<br>$a_c = 3.670\times10^{-7}$ m/s$^2$<br>$\lambda = 6.189\times10^{-5}$<br>$v_{esc} = 331.2$ m/s<br>$S_{1min} = 19.9$ km | $\sigma = 33.3$ kg/m$^2$<br>$a_c = 2.447\times10^{-7}$ m/s$^2$<br>$\lambda = 4.126\times10^{-5}$<br>$v_{esc} = 270.4$ m/s<br>$S_{1min} = 16.2$ km | $\sigma = 44.4$ kg/m$^2$<br>$a_c = 1.835\times10^{-7}$ m/s$^2$<br>$\lambda = 3.094\times10^{-5}$<br>$v_{esc} = 234.2$ m/s<br>$S_{1min} = 14.1$ km |
| $\mu = 0.6$<br>$P_r = 7.243\times10^{-6}$ N/m$^2$ | $\sigma = 22.2$ kg/m$^2$<br>$a_c = 3.263\times10^{-7}$ m/s$^2$<br>$\lambda = 5.502\times10^{-5}$<br>$v_{esc} = 312.3$ m/s<br>$S_{1min} = 18.7$ km | $\sigma = 33.3$ kg/m$^2$<br>$a_c = 2.175\times10^{-7}$ m/s$^2$<br>$\lambda = 3.668\times10^{-5}$<br>$v_{esc} = 254.9$ m/s<br>$S_{1min} = 15.3$ km | $\sigma = 44.4$ kg/m$^2$<br>$a_c = 1.631\times10^{-7}$ m/s$^2$<br>$\lambda = 2.750\times10^{-5}$<br>$v_{esc} = 220.8$ m/s<br>$S_{1min} = 13.2$ km |

**Trade Space for WISP Sprites 9 cm$^2$ area with 1.22 W Solar Energy Input**

Yet, instead, our plan is to deploy our sail system from a CubeSat, and then to push it with a 1-2 W laser, and then to capture it on film and video. The mission has several key objectives:

- Goal 1: Successful deployment of a sail from a CubeSat.
- Goal 2: Successful demonstration of laser-push of sail.
- Goal 3: Successful demonstration of a turning manoeuvre by moving the laser position by say 10° to effect a directional change.
- Goal 4: Provided the first three goals are achieved and the altitude high enough, our ambition is to successfully conduct one 90 minute orbit of the Earth.

An illustration of the i4is laser-sail concept is shown in the graphic below which shows a sail being propelled by a laser housed on the main CubeSat. For reliability it would be desirable to launch several and with different unfolding mechanisms and sail designs, so as to improve the chances that one will demonstrate proof of concept. The i4is is looking for direct sponsorship to fund this mission which we estimate is in the range $250,000+.

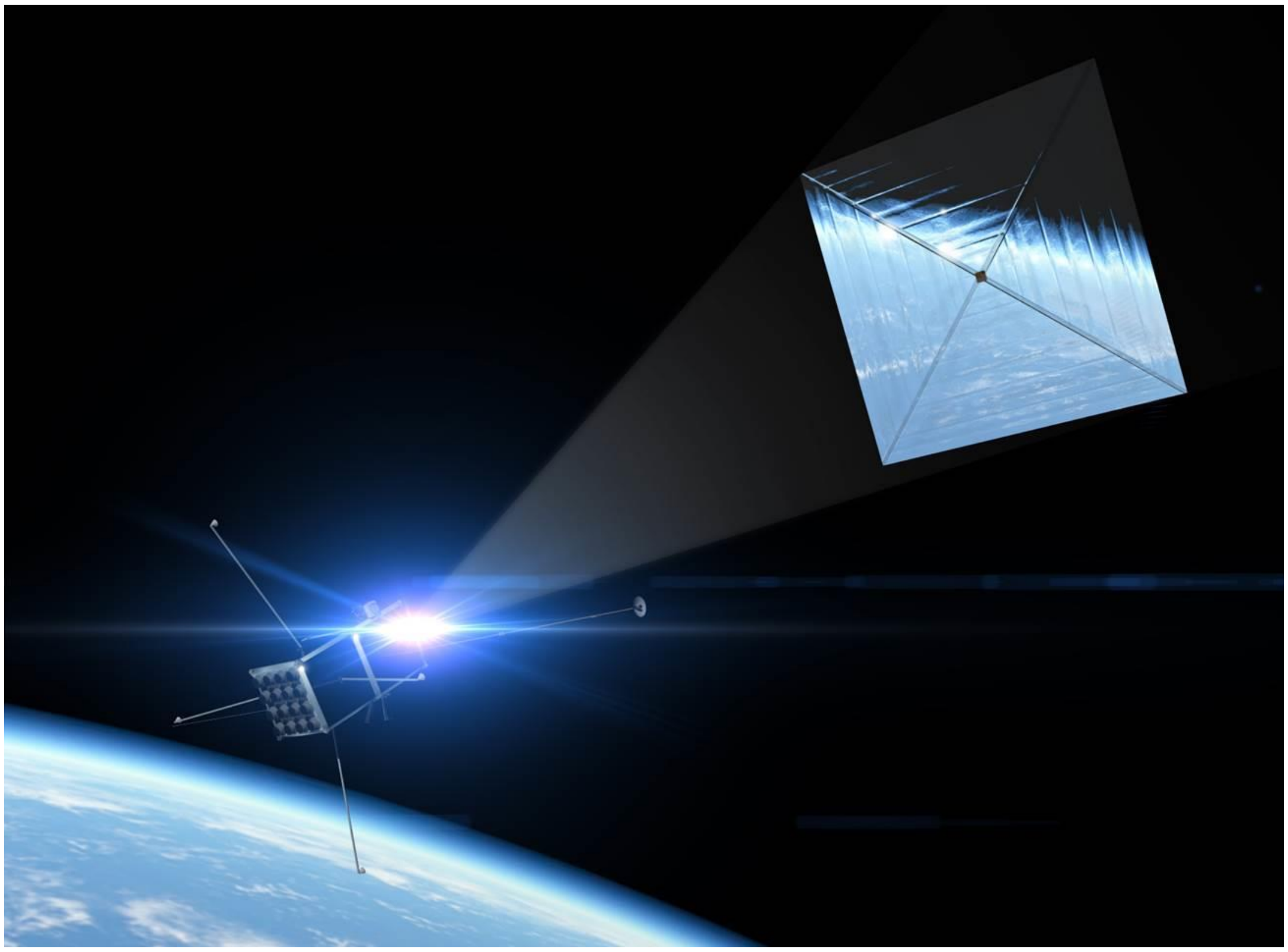

*Figure 19:* **Illustration of Laser-Sail CubeSat mission that i4is wants to conduct in 2017-2020 timeframe as a mission demonstrator**

The possible mission is summarised below.

*Mission statement*: The first in-space demonstration of laser sail propulsion

*Key mission requirements*: To increase the orbit semi-major axis of the laser sail spacecraft in low earth orbit (LEO) by at least 5 km via a laser beam from a CubeSat. 5 km are selected, as it is a measurable change in the orbit parameters.

*Mission architecture*: A CubeSat is launched into an 800 km orbit (lower orbits have higher drag, which requires a much bigger laser power to overcome it), Sun-synchronous, with a Right Ascension of the Ascending Node (RAAN) angle perpendicular to the Sun's position, as shown here:

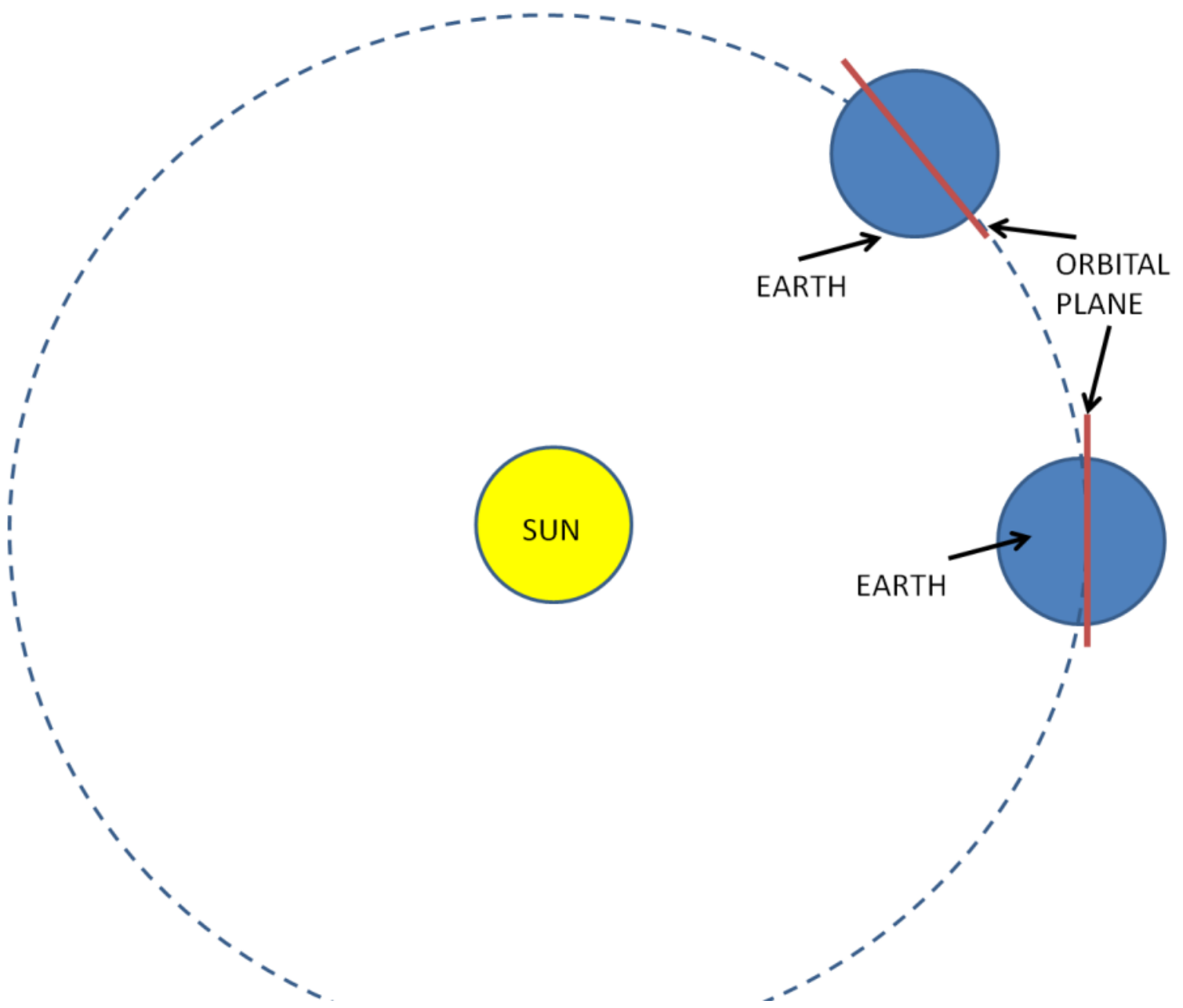

*Figure x. Schematic of orbit……..*

This has two advantages: the first is that the CubeSat, which has to carry the laser system, will almost always be in the view of the Sun, and therefore can get the required energy with the solar panels, without any eclipse time. The second is that, due to the perpendicular configuration, the Sun does not exert almost any force on the sail. The sail (20 cm x 20 cm) is deployed from the CubeSat. It has a total mass of 50 g (sail + ADCS + transmitter), as we need to send the position of the sail, and also actively control the sail attitude with some magnetic coils.

The CubeSat has an on-board electric micro-propulsion system, so it can chase the sail all the time and provide constant thrust, the distance from the sail is always kept constant, thereby having a constant pointing accuracy requirement. After 29 days of laser illumination, the required orbit increase of 5 km for the sail has been achieved.

*Maturity of used technologies*: All components have a TRL of 7 or higher. The CubeSat can be provided by FOTEC, Austria, a company for which i4is has links [56, 57, 58, 59].

*Estimated cost / schedule*: The cost can be estimated in less than 1 million US-$. Depending on the funding, this mission can be assembled in 18-24 months.

## 5. Minimal Interstellar Mission

In addition, our program included consideration for a minimal interstellar mission which will now briefly be described.

*Mission statement*: The first interstellar mission of Mankind (the Voyager probes are not directed to any particular star) and first private spacecraft to leave the Solar System. It intends to motivate future missions to overtake this spacecraft.

*Key mission requirements*: The objective of this mission is to beat the speed record of Voyager 1, the fastest space probe built by Mankind (**17 km/s**), arrive at Alpha Centauri, 4.24 light years away, in less than 25,000 years.

*Mission architecture*: Mission architecture consists of an interstellar probe, based on a 4U Cubesat, launched at solar system escape velocity by a suitable launcher. The probe is then propelled by a cluster of ion microthrusters, which are powered by an advanced Radioisotope Thermoelectric Generator.

Figure 1 shows an artist´s concept of the interstellar probe. The first Cubesat unit houses an array of ion microthrusters of the type Indium Field Emission Electric Propulsion. The Indium FEEP microthruster has the required specific impulse (4000 s) and the most efficient way of carrying propellant (Indium stored in solid state). Each thruster can provide a nominal thrust of 0.35 mN. A thrusting time of 20 years is needed for a **Δv of 25 km/s** which is enough to fulfill the flight time requirement of 25,000 years (as Alpha Centauri has a radial velocity of 25 km/s towards the Sun, the spacecraft velocity relative to AC will be 50 km/s). As the expected lifetime of an Indium FEEP microthruster is less than 5 years, a cluster of at least four or five microthrusters is needed to perform the mission. The Indium mass required for a thrusting time of 20 years is about 6 kg, which corresponds to a reservoir as large as a 1U CubeSat. Hence, the Indium FEEP propulsion system takes two CubeSats, leading to a 12 kg 4U interstellar probe (2 CubeSats for propulsion, 1 CubeSat for the power source and 1 CubeSat for the payload, see Table 1).

The propulsion system input power is about 30 W. The power source is an Advanced Radioisotope Thermoelectric Generator with an electrical power output per unit mass of 10 W/kg; this leads to a power source mass of at least 3 kg. The thermal power dissipated by the ARTG can be used to keep the thruster reservoir at an operational temperature above the Indium melting point.

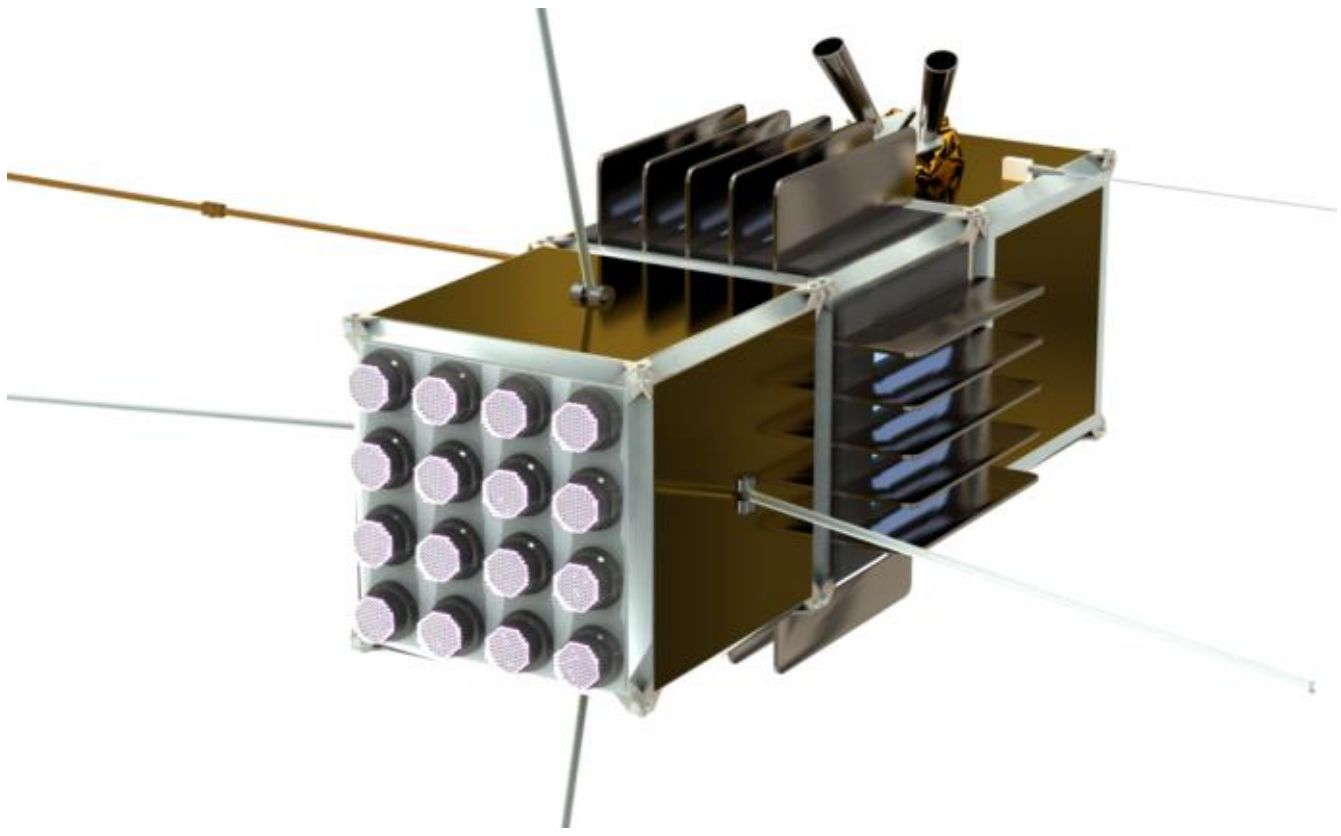

**Figure 1: an artist´s concept of the spacecraft (Credit: Adrian Mann); the first CubeSat houses an array of micro-thrusters, the second contains a miniaturized radioisotope thermal generator with affixed radiators, the third contains the payload. In this picture the 1U unit for the Indium propellant reservoir is missing.**

**Table 1: Preliminary mass budget**

| | |
|---|---|
| Propellant | 6 |
| Advanced RTG Power | 3 |
| Total Engine Assembly | 1 |
| Payload (instruments, comms…) | 1 |
| Spacecraft Structure | 1 |
| **Total Mass ~ 4U CubeSat** | **12 kg** |

*Alternatives and key parameter values for alternatives*: A ballistic launch with hyperbolic excess energy C3 of 250 $km^2/s^2$ could give a $\Delta v$ of 15 km/s in excess to the Solar System escape velocity. This launch energy should be achievable with a Falcon Heavy or SLS. A Jupiter gravity assist could add a $\Delta v$ of 10 km/s. Adding up these velocities with the 25 km/s provided by the ion microthrusters leads to a final speed relative to the Sun of **50 km/s (10 AU/year).** That is 500 AU in 50 years and Alpha Centauri in less than 20,000 years, accounting for the relative AC velocity.

If the advanced RTG is not available in time, the mission can be performed with a more standard RTG with a specific power of 5 W/kg, and compensating the lower $\Delta v$ provided by the ion propulsion system with a more energetic launch and/or a Jupiter gravity assist.

*Maturity of used technologies*:  The Indium FEEP microthrusters are under development at FOTEC, Austria, in the framework of an ESA contract (The i4is has links to FOTEC). The present TRL is 6 [60, 61, 62]. The advanced RTG is at a lower TRL.

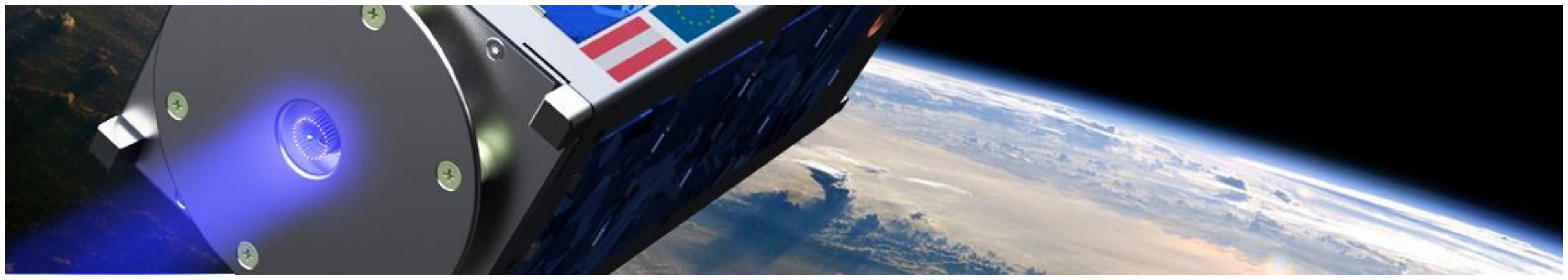

*Estimated cost / schedule***:** The cost of the interstellar probe can be estimated to be less than **5 million US-$.** Depending on the funding and on the availability of an advanced RTG, the interstellar probe can be developed in 3-5 years.

## 6. ANDROMEDA: ULTRALIGHT INTERSTLELAR FLYBY MISSION

Ultimately, our aim is to conduct a full interstellar mission and launch it within the next 20 years. We briefly describe what such a system may look like, and with a cut-down mass budget.

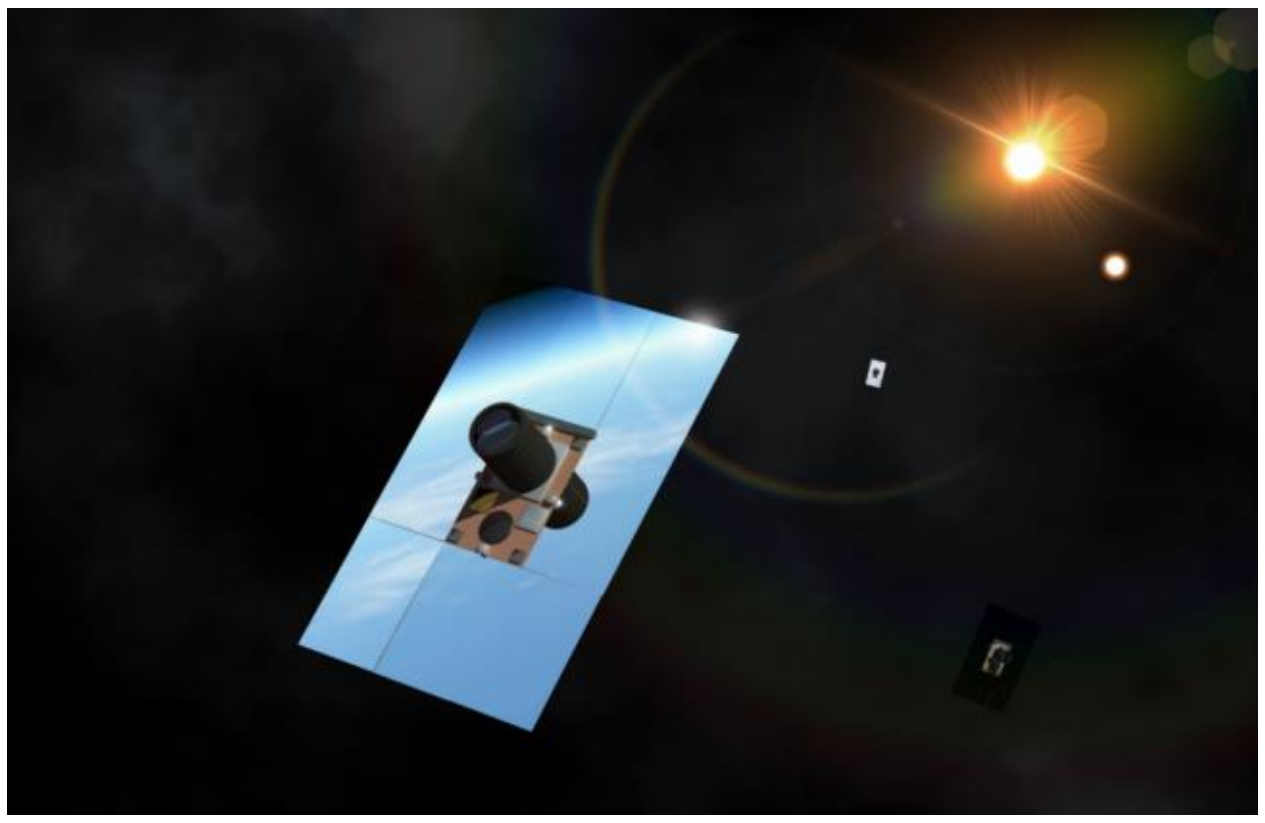

*Figure x. Interstellar mission*

*Mission statement*: To perform a fly-by mission to Alpha Centauri and return at least a photo of the star or potential exoplanets.

*Key mission requirements*:

- Maximum mission duration of 50 years

- Return at least one photo of the star and/or exoplanet
- Perform a fly-by mission

*Mission architecture*: Mission architecture consists of the interstellar probe and the space-based beaming infrastructure. The beaming infrastructure accelerates the probe via a laser or microwave beam. After acceleration, the probe enters the cruise phase. During target system encounter, pictures of the star and/or exoplanet are taken and sent back to the Solar System.

## Alternatives

| **Architecture element** | **Option 1** | **Option 2** | **Option 3** |
|---|---|---|---|
| Beaming infrastructure | Laser | Microwaves | |
| **Spacecraft** | | | |
| - *Power* | Electromagnetic tether + graphene capacitors | Nuclear battery + graphene capacitors | |
| - *Communication* | Radiofrequency | Laser | Star occultation |
| - *Navigation* | Pulsar / quasar navigation | Enhanced star tracker | |
| - *Interstellar dust protection* | Graphene whipple shield | Beryllium | |
| - *Radiation protection* | Radiation-tolerant electronic components | | |
| - *On-board data handling* | FPGA-based microprocessor | | |

Our preferred design is highlighted in green in the table above.

*Baseline architecture*: A laser infrastructure is selected as a baseline. In order to decrease the lens size, ten sequential Fresnel lenses with a radius of 95 m are used. Circular structures of this diameter are currently conceived by Tethers Unlimited for in-orbit manufacturing. A potential Fresnel lens material is Graphene sandwich. Graphene lenses have been demonstrated in the lab in 2016. The lens infrastructure is shown in the image below:

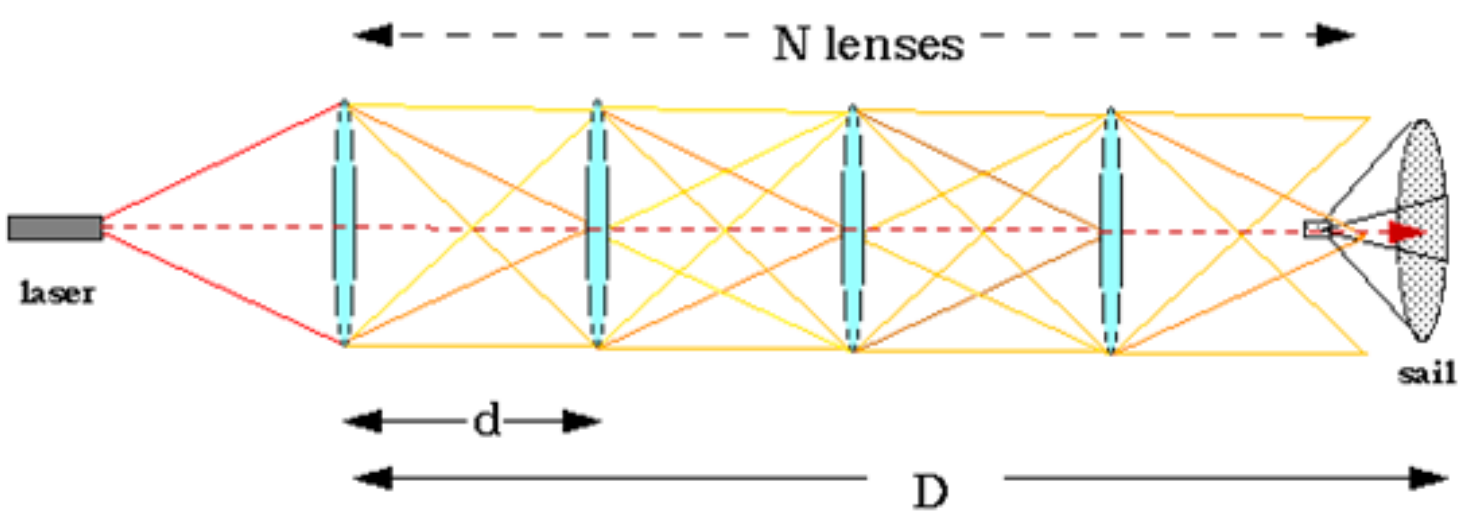

| | |
|---|---|
| Laser infrastructure power [MW] | 1150 |
| Lens radius [m] | 95 |
| Number of lenses | 10 |
| Acceleration distance [AU] | 1.8 |
| Cruise velocity [%c] | 10 |

***Table x. Estimated spacecraft mass budget***

| **Spacecraft subsystem** | **Technology** | **[g]** |
|---|---|---|
| Payload | MEMS camera + aperture, various MEMS sensors | 2,4 |
| OBDH | FPGA-based microcomputer | 1 |
| Power | Graphene supercapacitors (storage) and electromagnetic tethers (power generation) | 6.5 |
| Structure | Rigid Graphene matrix | 0,1 |
| Communications RF | Foldable phased-array antenna + transceiver | 3 |
| ADCS | Momentum wheels, MEMS FEEP thrusters | 1 |
| Navigation | Use of camera | - |
| Interstellar dust protection | Graphene Whipple shield | 2 |
| ***Bus mass*** | | ***15*** |
| Sail | 4-layer Graphene sandwich (Radius: 34m) | 8 |
| ***Total mass [g]*** | | ***23*** |

***Table x. Estimated cost***

Due to the miniaturization, common mass-based cost estimation models do not seem to be applicable to the spacecraft. We therefore select cost estimates for currently developed interplanetary CubeSats and multiply these costs by a factor of 5 in order to represent the increased difficulty of the mission. With a rough cost of 20 million dollar for an interplanetary CubeSat, we assume that developing the interstellar spacecraft may be comparable to that or several times higher in the 10s million US-$. The cost of the laser power generation infrastructure strongly depends on synergies between potential space power satellite systems. However, using near-term ultrathin solar cells with a specific power of 6 kg/kW we expect a cost of development, production, and launch to be in the billions US-$. The advantage of a space-based infrastructure are a factor 10 higher PV power output and laser efficiency compensating for high transportation cost.

# 7. CONCLUSIONS

In this report we have briefly considered the scenario of sending a small probe towards the Alpha Centauri system in a 50 year total mission time travelling at an average cruise velocity of 0.1c. We find that this idea is plausible and may be attempted within the next 10-20 years provided adequate investment is made and also that key laser technology is permitted into the space environment. The specific probe design down selected has a total spacecraft mass of 280 grams propelled by a 1.1 GW 500nm beam, or a smaller design at 28 grams in mass.

On a final note, the *Initiative for Interstellar Studies* team would like to conduct a space mission attempt during the 2017-2020 time-frame which demonstrates the first laser-sail in space using ChipSat technology. We estimate that such a mission cost would be in the range of $250,000+ and this concept is directly on the technology roadmap towards the grander concept described in this report. We are looking for funding to support this attempt as a way of kickstarting global efforts towards laser-sail propulsion. Our philosophy for this mission is to keep it simple, cheap, do it quickly and to minimise risk by ground validation and physics demonstration. Our team is ready to go and is looking for sponsorship.

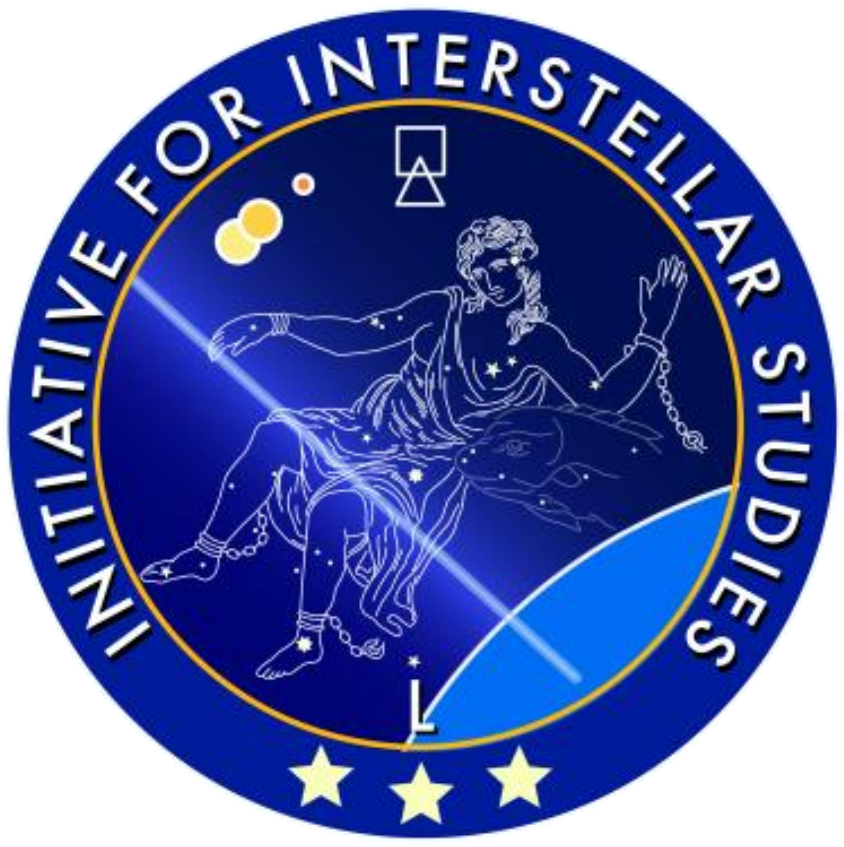

*Figure x. Project Andromeda mission patch*

## ACKNOWLEDGEMENTS

In addition to the Interstellar Probe Team the following are thanked for their contributions to this study: Gregory Matloff, Peter Milne, Sam Harrison.

Companies, 1999.